\documentclass[11pt]{article}
\usepackage{amsmath}
\usepackage{amssymb}
\usepackage{amsfonts}
\usepackage{amsthm}
\usepackage[utf8]{inputenc}
\usepackage{changepage}
\usepackage{float} 
\usepackage{booktabs,caption,fixltx2e}
\usepackage[flushleft]{threeparttable}
\usepackage{graphicx}
\usepackage{footnote}
\usepackage{caption}
\usepackage{subcaption}
\usepackage{geometry}
\usepackage{graphicx}
\usepackage[toc,page]{appendix}
\usepackage{chngcntr}
\usepackage{color, colortbl}
\usepackage{rotating}
\usepackage[british]{babel}
\usepackage{times} 
\usepackage{color}
\usepackage{mathtools}
\usepackage{empheq}
\usepackage{graphicx}
\usepackage{threeparttable}
\usepackage{amsmath,amsfonts,amssymb,amsthm,mathtools}
\usepackage{chngcntr}
\usepackage{pgfplots}
\usepackage{pgfplotstable}
\usepgfplotslibrary{dateplot} 
\pgfplotsset{compat=newest}
\usepackage{remreset}
\usepackage{bm}
\usepackage{tikz} 
\usepackage[round]{natbib} 
\usepackage{authblk}
\usepackage{soul} 
\usepackage[normalem]{ulem} 
\usepackage[font=small]{caption}
\usepackage{hyperref}

\definecolor{lgray}{gray}{0.8}

\newcommand{\newoperator}[3]{\newcommand*{#1}{\mathop{#2}#3}}
\newcommand{\renewoperator}[3]{\renewcommand*{#1}{\mathop{#2}#3}}
\newcommand{\mB}{\bm B}

\newcommand{\mC}{\bm C}

\newcommand{\ve}{\bm e}
\newcommand{\mF}{\bm F}
\newcommand{\vf}{\bm f}

\newcommand{\mH}{\bm H}

\newcommand{\mI}{\bm I}

\newcommand{\mP}{\bm P}

\newcommand{\mQ}{\bm Q}

\newcommand{\mR}{\bm R}

\newcommand{\mT}{\bm T}

\newcommand{\vu}{\bm u}

\newcommand{\vv}{\bm v}

\newcommand{\vx}{\bm x}

\newcommand{\vy}{\bm y}
\newcommand{\mZ}{\bm Z}

\newcommand{\valpha}{\bm \alpha}
\newcommand{\vbeta}{\bm \beta}

\newcommand{\vvarepsilon}{\bm \varepsilon}

\newcommand{\veta}{\bm \eta}

\newcommand{\vlambda}{\bm \lambda}

\newcommand{\vxi}{\bm \xi}

\newcommand{\vrho}{\bm \rho}

\newcommand{\mLambda}{\bm \varLambda}

\newcommand{\mPsi}{\bm \varPsi}

\renewoperator{\Re}{\mathrm{Re}}{\nolimits}
\renewoperator{\Im}{\mathrm{Im}}{\nolimits}
\makeatletter
\newcommand{\rd}{\@ifnextchar^{\DIfF}{\DIfF^{}}}
\def\DIfF^#1{%
   \mathop{\mathrm{\mathstrut d}}%
   \nolimits^{#1}\gobblespace}
\def\gobblespace{\futurelet\diffarg\opspace}
\def\opspace{%
   \let\DiffSpace\!%
   \ifx\diffarg(%
   \let\DiffSpace\relax
   \else
   \ifx\diffarg[%
   \let\DiffSpace\relax
   \else
   \ifx\diffarg\{%
   \let\DiffSpace\relax
   \fi\fi\fi\DiffSpace}

\newcommand{\bias}{\operatorname{bias}}

\newcommand{\cov}{\operatorname{cov}}

\newcommand{\diag}{\operatorname{diag}}
\newcommand{\blockdiag}{\operatorname{blockdiag}}

\newoperator{\ip}{\mathrm{int}}{\nolimits}

\newcommand{\MSE}{\operatorname{MSE}}
\newcommand{\MSFE}{\operatorname{MSFE}}

\newcommand{\var}{\operatorname{var}}

\newcommand{\vones}{\bm\imath}

\newcommand{\vzeros}{\boldsymbol{0}}

\numberwithin{equation}{section}

\theoremstyle{plain}

\pgfplotsset{select coords between index/.style 2 args={
    x filter/.code={
        \ifnum\coordindex<#1\fi
        \ifnum\coordindex>#2\fi
    }
}}

\title{A dynamic factor model approach to incorporate Big Data in state space models for official statistics\thanks{This work was funded by the European Union under grant no. 07131.2017.003-2017.596. The views expressed in this paper are those of the authors and do not necessarily reflect the policy of Statistics Netherlands. 
Previous versions of this paper have been presented at CFE-CM Statistics 2017, The Netherlands Econometric Study Group 2018, Small Area Estimation 2018, Methods for Big Data in Official Statistics, BigSurv 2018, the 29th $(\text{EC})^2$ on Big Data Econometrics with Applications, and at internal seminars organized by Maastricht University and Statistics Netherlands. We thank conference and seminar participants for their interesting comments. Additionally, we thank Marco Puts and Ole Mussmann for their help with the data collection. All remaining errors are our own.
}}

\author[1,2]{Caterina Schiavoni\thanks{Corresponding author: Department of Quantitative Economics, Maastricht University, P.O. Box 616, 6200 MD Maastricht, The Netherlands. E-mail: c.schiavoni@maastrichtuniversity.nl.}}
\author[2]{Franz Palm}
\author[2]{Stephan Smeekes}
\author[1,2]{Jan van den Brakel}
\affil[1]{Statistics Netherlands, Heerlen, The Netherlands}
\affil[2]{Department of Quantitative Economics, Maastricht University, The Netherlands}

\begin{document}


\maketitle


\begin{abstract}
In this paper we consider estimation of unobserved components in state space models using a dynamic factor approach to incorporate auxiliary information from high-dimensional data sources. 
We apply the methodology to unemployment estimation as done by Statistics Netherlands, who uses a multivariate state space model to produce monthly figures for the unemployment using series observed with the labour force survey (LFS). We extend the model by including auxiliary series of Google Trends about job-search and economic uncertainty, and claimant counts, partially observed at higher frequencies. Our factor model allows for nowcasting the variable of interest, providing reliable unemployment estimates in real-time before LFS data become available. 
\newline 

{\bf Keywords:}  high-dimensional data analysis, state space, factor models, nowcasting, unemployment, Google Trends.

\end{abstract}


\section{Introduction}
There is an increasing interest among national statistical institutes (NSIs) to use data that are generated as a by-product of processes not directly related to statistical production purposes in the production of official statistics. Such data sources are sometimes referred to as ``Big Data''; examples are time and location of network activity available from mobile phone companies, social media messages from Twitter and Facebook, sensor data, and internet search behaviour from Google Trends.
A common problem with this type of data sources is that they are likely selective with respect to an intended target population. If such data sources are directly used to produce statistical information, then the potential selection bias of these data sources must be accounted for, which is often a hard task since Big Data sources are often noisy and generally contain no auxiliary variables, which are required for bias correction. These problems can be circumvented by using them as covariates in model-based inference procedures to make precise detailed and timely survey estimates, since they come at a high frequency and are therefore very timely. These techniques are known in the literature as small area estimation and nowcasting \citep{RaoMolina2015}.

Official statistics are generally based on repeated samples. Therefore multivariate time series models are potentially fruitful to improve the precision and timeliness of domain estimates with survey data obtained in preceding reference periods and other domains. The predictive power of these models can be further improved by incorporating auxiliary series that are related with the target series observed with a repeated survey.

In this paper we investigate how auxiliary series derived from big data sources and registers can be combined with time series observed with repeated samples in high dimensional multivariate structural time series (STS) models. We consider Google Trends and claimant counts as auxiliary series for monthly unemployment estimates observed with a continuously conducted sample survey. Big Data sources have the problem that they are noisy and potentially (partly) irrelevant, and, as such, care must be taken when using them for the production of official statistics. We show that, by using a dynamic factor model in state space form, relevant information can be extracted from such auxiliary high-dimensional data sources, while guarding against the inclusion of irrelevant data. 



Statistical information about a country's labour force is generally obtained from labour force surveys, since the required information is not available from registrations or other administrative data sources. The Dutch labour force survey (LFS) is based on a rotating panel design, where monthly household samples are observed five times with quarterly intervals. These figures are, however, considered too volatile to produce sufficiently reliable monthly estimates for the employed and the unemployed labour force at monthly frequency. For this reason Statistics Netherlands estimates monthly unemployment figures, together with its change, as unobserved components in a state space model where the observed series come from the monthly Dutch LFS, using a model originally proposed by \cite{Pfeffermann1991}. This method improves the precision of the monthly estimates for unemployment with sample information from previous periods, and can therefore be seen as a form of small area estimation. In addition it accounts for rotation group bias \citep{bailar1975}, serial correlation due to partial sample overlap, and discontinuities due to several major survey redesigns \citep{VanDenBrakel2015}.

Time series estimates for the unemployment can be further improved by including related auxiliary series. The purpose is twofold. First, auxiliary series can further improve the precision of the time series predictions. In this regard, \cite{harvey2000} propose a bivariate state space model to combine a univariate series of the monthly unemployed labour force derived from the UK LFS, with the univariate auxiliary series of claimant counts. The latter series represents the number of people claiming unemployment benefits. It is an administrative source, which is not available for every country, and, as for the Netherlands, it can be affected by the same publication delay of the labour force series. Second, auxiliary series derived from Big Data sources like Google Trends are generally available at a higher frequency than the monthly series of the LFS. Combining both series in a time series model allows to make early predictions for the survey outcomes in real-time at the moment that the outcomes for the auxiliary series are available, but the survey data not yet, which is in the literature known as \emph{nowcasting}, in other words, ``forecasting the present".

In this paper, we extend the state space model used by Statistics Netherlands in order to combine the survey data with the claimant counts and the high-dimensional auxiliary series of Google Trends about job-search and economic uncertainty, as they could yield more information than a univariate one, which is not affected by publication lags and that can eventually be observed at a higher frequency than the labour force series. 

This paper contributes to the existing literature by proposing a method to include a high-dimensional auxiliary series in a state space model in order to improve the (real-time) estimation of unobserved components. The model accounts for the rotating panel design underlying the sample survey series, combines series observed at different frequencies, and deals with missing observations at the end of the sample due to publication delays. It handles the curse of dimensionality that arises from including a large number of series related to the unobserved components, by extracting their common factors.

Besides claimant counts, the majority of the information related to unemployment is nowadays available on the internet; from job advertisements to resum\'e's templates and websites of recruitment agencies. We therefore follow the idea originating in \cite{Choi2009}, \cite{askitas2009} and \cite{Suhoy2009} of using terms related to job and economic uncertainty, searched on Google in the Netherlands. Since 2004, these time series are freely downloadable in real-time from the Google Trends tool, on a monthly or higher frequency. As from the onset it is unclear which search terms are relevant, and if so, to which extent, care must be taken not to model spurious relationships with regards to the labour force series of interest, which could have a detrimental effect on the estimation of unemployment, such as happened for the widely publicized case of Google Flu Trends \citep{Lazer2014}.

Our method allows to exploit the high-frequency and/or real-time information of the auxiliary series, and to use it in order to nowcast the unemployment, before the publication of labour force data. As the number of search terms related to unemployment can easily become large, we employ the two-step estimator of \cite{Doz2011}, which combines factor models with the Kalman filter, to deal both with the high-dimensionality of the auxiliary series, and with the estimation of the state space model. The above-mentioned estimator is generally used to improve the nowcast of variables that are observed such as GDP (see \cite{giannone2008} and \cite{Hindrayanto2016} for applications to the US and the euro area), which is not the case for the unemployment. Nonetheless, \cite{DAmuriMarcucci2017}, \cite{Naccaratoetal2018}, and \cite{Maas2019} are all recent studies that use Google Trends to nowcast and forecast the unemployment, by treating the latter as \emph{known} dependent variable in time series models where the Google searches are part of the explanatory variables. To the best of our knowledge, our paper is the first one to use Google Trends in order to nowcast the unemployment in a model setting that treats the latter variable as \emph{unobserved}.

We evaluate the performance of our proposed method via Monte Carlo simulations and find that our method can yield large improvements in terms of Mean Squared Forecast Error $(\MSFE)$ of the unobserved components' nowcasts. We then assess whether the accuracy of the unemployment's estimation and nowcast improves with our high-dimensional state space model, respectively from in-sample and out-of-sample results. The latter consists of a recursive nowcast. We do not venture into forecasting exercises as Google Trends are considered to be more helpful in predicting the present rather than the future of economic activities \citep{ChoiVarian2012}. We conclude that Google Trends can significantly improve the fit of the model, although the magnitude of these improvements is sensitive to aspects of the data and the model specification, such as the frequency of observation of the Google Trends, the number of Google Trends' factors included in the model, and the level of estimation accuracy provided by the first step of the two-step estimation procedure.

The remainder of the paper is organized as follows. Section \ref{DataLFS} discusses the data used in the empirical analysis. Section \ref{DLFS} describes the state space model that is currently used by Statistics Netherlands to estimate the unemployment. Section \ref{auxiliary} focuses on our proposed method to include a high-dimensional auxiliary series in the aforementioned model. Sections \ref{Simulations} and \ref{empirics} report, respectively, the simulation and empirical results for our method. Section \ref{conclusions} concludes.


\section{Data}  \label{DataLFS}


The Dutch LFS is conducted as follows. Each month a stratified two-stage cluster design of addresses is selected. Strata are formed by geographical regions. Municipalities are considered as primary sampling units and addresses as secondary sampling units. All households residing on an address are included in the sample with a maximum of three (in the Netherlands there is generally one household per address). All household members with age of 16 or older are interviewed. Since October 1999, the LFS has been conducted as a rotating panel design. Each month a new sample, drawn according to the above-mentioned design, enters the panel and is interviewed five times at quarterly intervals. The sample that is interviewed for the $j^{th}$ time is called the $j^{th}$ wave of the panel, $j=1,\dots,5$. After the fifth interview, the sample of households leaves the panel. This rotation design implies that in each month five independent samples are observed. The generalized regression (GREG, i.e., design-based) estimator \citep{sarndal1992} is used to obtain five independent direct estimates for the unemployed labour force, which is defined as a population total. This generates over time a five-dimensional time series of the unemployed labour force. Table \ref{tab:visualization} provides a visualization for the rotation panel design of the Dutch LFS.

\begin{center}
\begin{tabular}{c c c c c c r}
$\overbracket{\overbracket{\text{A}}^{\text{month}} \text{B C}}^{\text{quarter}}$ & D E F & G H I & J K L & M N O & P Q R &  $ \begin{rcases} \end{rcases} \text{\scriptsize{wave 1}}$ \\ 
 & A B C & D E F & G H I & J K L & M N O & $ \begin{rcases} \end{rcases} \text{\scriptsize{wave 2}}$ \\  
 & & A B C & D E F & G H I & J K L & $ \begin{rcases} \end{rcases} \text{\scriptsize{wave 3}}$ \\ 
& & & A B C & D E F & G H I & $ \begin{rcases} \end{rcases} \text{\scriptsize{wave 4}}$ \\
& & & & A B C & D E F & $ \begin{rcases} \end{rcases} \text{\scriptsize{wave 5}}$ \\
\end{tabular} 
\captionof{table}{Visualization for the rotation panel design of the Dutch LFS. Each capital letter represents a sample. Every month a new sample enters the panel and is interviewed five times at a quarterly frequency. After the fifth interview, the sample of households leaves the panel.} \label{tab:visualization}
\end{center}

Rotating panel designs generally suffer from Rotation Group Bias (RGB), which refers to the phenomena that there are systematic differences among the observations in the subsequent waves \citep{bailar1975}. In the Dutch LFS the estimates for the unemployment based on the first wave are indeed systematically larger compared to the estimates based on the follow-up waves \citep{VanDenBrakel2015}. This is the net results of different factors:
\begin{itemize}
\item Selective nonresponse among the subsequent waves, i.e., panel attrition.
\item Systematic differences due to different data collection models that are applied to the waves. Until 2010 data collection in the first wave was based on face-to-face interviewing. Between 2010 and 2012 data collection in the first wave was based on telephone interviewing for households for which a telephone number of a landline telephone connection was available and face-to-face interviewing for the remaining households. After 2012 data collection in the first wave was based on a sequential mixed mode design that starts with Web interviewing with a follow up using telephone interviewing and face-to-face interviewing. Data collection in the follow-up waves is based on telephone interviewing only.
\item Differences in wording and questionnaire design used in the waves. In the first wave a block of questions is used to verify the status of the respondent on the labour force market. In the follow-up waves the questionnaire focuses on differences that occurred compared to the previous interview, instead of repeating the battery of questions.
\item Panel conditioning effects, i.e., systematic changes in the behaviour of the respondents. For example, questions about activities to find a job in the first wave might increase the search activities of the unemployed respondents in the panel. Respondents might also systematically adjust their answers in the follow-up waves, since they learn how to keep the routing through the questionnaire as short as possible.
\end{itemize}

The Dutch labour force is subject to a one-month publication delay, which means that the sample estimates for month $t$ become available in month $t+1$. In order to have more timely and precise estimates of the unemployment, we extend the model by including, respectively, auxiliary series of weekly/monthly Google Trends about job-search and economic uncertainty, and monthly claimant counts, in the Netherlands.

Claimant counts are the number of registered people that receive unemployment benefits. The claimant counts for month $t$ become available in month $t+1$.

Google Trends are indexes of search activity. Each index measures the fraction of queries that include the term in question in the chosen geography at a particular time, relative to the total number of queries at that time. The maximum value of the index is set to be 100. According to the length of the selected period, the data can be downloaded at either monthly, weekly, or higher frequencies. The series are standardized according to the chosen period and their values can therefore vary according to the period's length \citep{Stephens-Davidowitz2015a}. 
We use weekly and monthly Google Trends for each search term. Google Trends are available in real-time (i.e., they are available in period $t$ for period $t$, independently on whether the period is a week or a month).

The list of Google search terms used in the empirical analysis of this paper, together with their translation/explanation, is reported in Tables \ref{tab:google_searches1} and \ref{tab:google_searches2}. A first set of terms (which is the one used in a previous version of this paper) was chosen by thinking of queries that could be made by unemployed people in the Netherlands. The rest of the terms has been chosen by using the Google Correlate tool and selecting the queries that are highly correlated to each term of the initial set, and that have a meaningful relation to unemployment and, more generally, economic uncertainty\footnote{Later in the paper we mention that we need non-stationary (e.g., persistent) Google Trends for our model. Correlations between non-stationary series can be spurious, and in this respect Google Correlate is not an ideal tool in order to choose search terms. In section \ref{empirics} we explain how to circumvent this problem.}.

\begin{figure}[h!]
\centering
\begin{tikzpicture}
\begin{axis}[axis lines*=left,height=4.5cm, width=0.9\linewidth, date coordinates in=x, table/col sep=comma, date ZERO=2004-01-01, xticklabel=\year-\month, xticklabel style={rotate=30, anchor=near xticklabel},legend style={at={(0.3,-0.5)},anchor=north,legend columns=2}, ylabel=Number of people]
\addplot[black] table[x index=0,y index=1] {waves.csv};
\addplot[black] table[x index=0,y index=2] {waves.csv};
\addplot[black] table[x index=0,y index=3] {waves.csv};
\addplot[black] table[x index=0,y index=4] {waves.csv};
\addplot[black] table[x index=0,y index=5] {waves.csv};
\addplot+[color=green, no markers, solid] table[x index=0,y index=6] {waves.csv};
\legend{,,,,$\vy^k_t$,claimant counts};
\end{axis}
\begin{axis}[height=4.5cm,
width=0.9\linewidth, date coordinates in=x, table/col sep=comma, date ZERO=2004-01-01, xticklabel=\year-\month, xticklabel style={rotate=30, anchor=near xticklabel},legend style={at={(0.7,-0.5)},anchor=north,legend columns=2},axis y line*=right, hide x axis, ylabel= Google trend index]
\addplot+[color=red, no markers, solid] table[x index=0,y index=7] {waves.csv};
\legend{``werkloos"};
\end{axis}
\end{tikzpicture}
\caption{Monthly time series of the five waves of the Dutch unemployed labour force ($\vy^k_t$), the claimant counts, and the Google search term ``werkloos", which means ``unemployed", in the Netherlands. The period starts in January 2004 and ends in May 2019.} \label{fig:timeseries}
\end{figure}
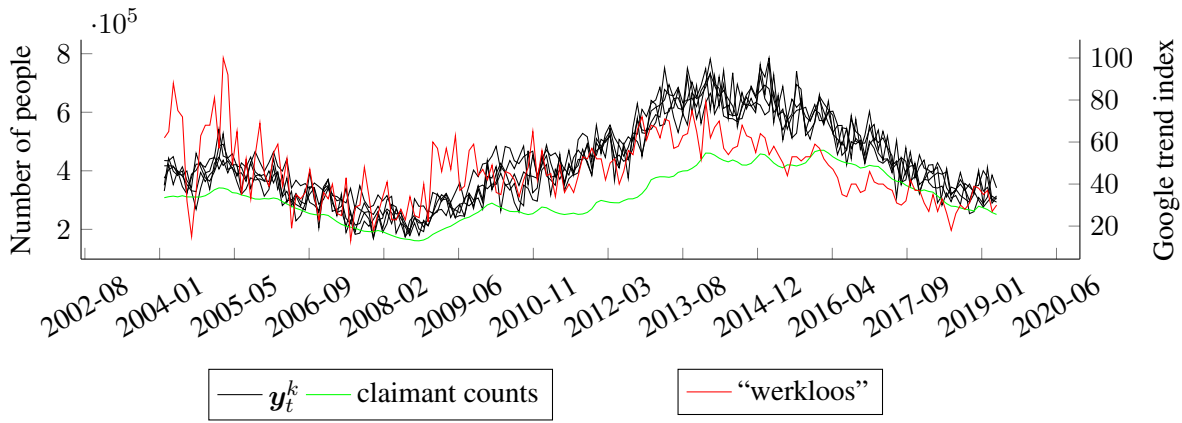

Figure \ref{fig:timeseries} displays the time series of the five waves of the unemployed labour force, together with the claimant counts and an example of job-related Google query. They all seem to be following the same trend, which already shows the potential of using this auxiliary information in estimating the unemployment.

\section{The Dutch labour force model and extensions} 

We first describe the model in use at Statistics Netherlands in Section 3.1. Next we explain how high-dimensional auxiliary series can be added to this  model in Section 3.2. 


\subsection{The Dutch labour force model} \label{DLFS}

The monthly sample size of the Dutch LFS is too small to produce sufficiently precise estimates directly. In the past, rolling quarterly figures were published on a monthly frequency. This has the obvious drawback that published figures are unnecessarily delayed since the reference period is the mid month of the rolling quarter. Also, real monthly seasonal effects are smoothed over the rolling quarter. Another problem that arose after the change from a cross-sectional survey to a rotating panel design in 2000, was that the effects of RGB became visible in the labour force figures. Both problems are solved with a structural time series (STS) model, that is used by
Statistics Netherlands, since 2010, for the production of monthly statistics about the Dutch labour force \citep{VanDenBrakel2015}. In a STS model, an observed series is decomposed in several unobserved components, such as a trend, a seasonal component, one or more cycles with a period longer than one year, regression components, and a white noise component. After writing an STS model in the state space form, the Kalman filter can be applied in order to estimate the unobserved components. See \citet{durbinkoopman2012} for an introduction to STS modelling.

Let $y^k_{j,t}$ denote the GREG estimate for the unemployment in month $t$ based on the sample observed in wave $j$. Now $\vy^k_t = (y^k_{1,t}, \dots, y^k_{5,t})$ denotes the vector with the five GREG estimates for the unemployment in month $t$. The $y^k_{j,t}$ are treated as five distinct time series in a five dimensional time series model in order to account for the rotation group bias. The superscript $k > 1$ indicates that the vector is observed at the low frequency. We need this notation \citep[see e.g.][]{banbura2013} to distinguish between series observed at different frequencies, because later on we will make use of Google Trends which are available on a weekly basis. If $\vy^k_t$ is observed at the monthly frequency, as in the case of the unemployed labour force, then $k=4,5$ if the high frequency series is observed at the weekly frequency, since a month can have either 4 or 5 weeks. 

The unemployment is estimated, with the Kalman filter, as a state variable in a state space model where $\vy^k_t$ represents the observed series. The measurement equation takes the form
\citep{Pfeffermann1991, Brakel2009}:
\begin{equation} \label{eq:univ_measurement}
\vy^k_t = \vones_5 \theta^{k,y}_t + \vlambda^k_t  + \ve^k_t.
\end{equation}
where $\vones_5$ is a 5-dimensional vector of ones, and $\theta^{k,y}_t$, i.e. the unemployment, is the common population parameter among the five-dimensional waves of the unemployed labour force. It is composed of the level of a trend ($L_t$) and a seasonal component ($S_t$):
\begin{equation*}
 \theta^{k,y}_t  = L^{k,y}_t +   S^{k,y}_t.
\end{equation*}
The transition equations for the level ($L_t$) and the slope ($R_t$) of the trend are, respectively:
\begin{equation*}
\begin{split}
L^{k,y}_t  &= L^{k,y}_{t-1} + R^{k,y}_{t-1} , \\
R^{k,y}_t  &= R^{k,y}_{t-1}  + \eta^{k,y}_{R,t} , \quad \eta^{k,y}_{R,t}  \sim N\left( 0 , \sigma^2_{R,y} \right),
\end{split}
\end{equation*}
which characterize a smooth trend model. 
This implies that the level of the trend is integrated of order 2, denoted as $I(2)$, which means that the series of the level is stationary (i.e., mean-reverting) after taking two times successive differences. The slope of the trend, $R^{k,y}_t$, is a first-order integrated series, denoted as $I(1)$. This state variable represents the change in the level of the trend, $L^{k,y}_t$, and not in the unemployment, $\theta^{k,y}_t$, directly. Nevertheless, since the $I(2)$ property of the unemployment is driven by its trend, and not by its seasonal component, the change in $\theta^{k,y}_t$ will also mainly be captured by $R^{k,y}_t$, and we can therefore consider the latter as a proxy for the change in unemployment. The model originally contained an innovation term for the population parameter $\theta^{k,y}_t$. However, the maximum likelihood estimate for its variance tended to be zero and \cite{bollineni2017} showed via simulations that it is better to not include this term in the model.

The trigonometric stochastic seasonal component allows for the seasonality to vary over time, and it is modeled as in \citet[Chapter~3]{durbinkoopman2012}: \begin{equation*}
\begin{split}
 S^{k,y}_t &= \sum_{l=1}^6 S^{k,y}_{l,t}, \\
\left( \begin{array}{c} 
S^{k,y}_{l,t}\\
S^{*k,y}_{l,t}
\end{array}\right) &= \left[ \begin{array}{cc} 
\cos(h_l) &  \sin(h_l)\\
-\sin(h_l) & \cos(h_l)
\end{array}\right] \left( \begin{array}{c} 
S^{k,y}_{l,t-1}\\
 S^{*k,y}_{l,t-1}
\end{array}\right) + \left( \begin{array}{c} 
\eta^{k,y}_{\omega,l,t}\\
\eta^{*k,y}_{\omega,l,t}
\end{array}\right), \quad  \left( \begin{array}{c} 
\eta^{k,y}_{\omega,l,t} \\
 \eta^{*k,y}_{\omega,l,t} 
\end{array}\right) \sim N \left( \vzeros , 
\sigma^2_{\omega,y} \mI_2 \right), 
\end{split}
\end{equation*}
where $h_l = \frac{\pi l}{6}$, for $l=1,\dots,6$. 

The second component in equation \eqref{eq:univ_measurement}, $ \vlambda^k_t = (\lambda^k_{1,t}, \ldots, \lambda^k_{5,t})^t$, accounts for the RGB. Based on the factors that contribute to the RGB, as mentioned in Section \ref{DataLFS}, the response observed in the first wave is assumed to be the most reliable one and not to be affected by the RGB \citep{Brakel2009}. Therefore it is assumed that $\lambda^k_{1,t}  = 0$. The remaining four components in $\vlambda^k_t$ are random walks that capture time-dependent differences between the follow-up waves with respect to the first wave:
\begin{equation*}
\begin{split}
\lambda^k_{1,t} & = 0, \\
\lambda^k_{j,t} & = \lambda^k_{j,t-1} + \eta^k_{\lambda,j,t}, \quad \eta^k_{\lambda,j,t} \sim N \left( 0 , \sigma^2_{\lambda} \right), \quad j =2,\dots,5.
\end{split}
\end{equation*} 
As a result the Kalman filter estimates for $\theta^{k,y}_t $ in \eqref{eq:univ_measurement} are benchmarked to the level of the GREG series of the first wave.

The third component in equation \eqref{eq:univ_measurement}, $ \ve^k_t = (e^k_{1,t}, \ldots, e^k_{5,t})^t$, models
the autocorrelation among the survey errors ($e^k_{j,t}$) in the follow-up waves due to the sample overlap of the rotating panel design. In order to account for this autocorrelation, the survey errors are treated as state variables, which follow the transition equation below. 
\begin{equation} \label{eq:univ_sampling_error}
\begin{split}
& e^k_{j,t}= c_{j,t} \tilde{e}^k_{j,t}, \quad c_{j,t} = \sqrt{ \widehat{ \var} \left(y^k_{j,t}\right)}, \quad j=1,\dots,5, \\
& \tilde{e}^k_{1,t} \sim N \left(0,\sigma^2_{\nu_{1}}\right), \\
& \tilde{e}^k_{j,t} = \delta \tilde{e}^k_{{j-1},{t-3}} + \nu^k_{j,t}, \quad
\nu^k_{j,t} \sim N \left( 0 , \sigma^2_{\nu_{j}}\right), \quad j =2,\dots,5, \quad |\delta| < 1. \\
& \var\left(\tilde{e}^k_{j,t}\right) = \sigma^2_{\nu_{j}}/\left(1-\delta^2\right), \quad  j=2,\dots,5,
\end{split}
\end{equation}
with $\widehat{ \var} \left(y^k_{j,t}\right)$ being the design variance of the GREG estimates $y^k_{j,t}$.
The scaled sampling errors, $\tilde{e}^k_{j,t}$, for $j=1,\dots,5$, account for the serial autocorrelation induced by the sampling overlap of the rotating panel. Samples in the first wave are observed for the first time and therefore its survey errors are not autocorrelated with survey errors of previous periods. The survey errors of the second to fifth wave are correlated with the survey errors of the previous wave three months before. 
Based on the approach proposed by \cite{Pfeffermann1998}, \cite{Brakel2009} motivate that these survey errors should be modelled as an AR(3) process, without including the first two lags. Moreover, the survey errors of all waves are assumed to be proportional to the standard error of the GREG estimates. In this way the model accounts for heterogeneity in the variances of the survey errors, which are caused by changing sample sizes over time. As a result the maximum likelihood estimates of the variances of the scaled sampling errors, $\sigma^2_{\nu_{j}}$, will have values approximately equal to one.

The structural time series model \eqref{eq:univ_measurement} as well as the models proposed in the following sections are fitted with the Kalman filter after putting the model in state space form. 
We use an exact initialization for the initial values of the state variables of the sampling error, and a diffuse initialization for the other state variables.
It is common to call \textit{hyperparameters} the parameters that define the stochastic properties of the measurement equation and the transition equation of the state space model. These are the parameters that are assumed to be known in the Kalman filter
\citep[Chapter~2]{durbinkoopman2012}. In our case the hyperparameters are $\delta$ and all the parameters that enter the covariance matrices of the innovations. These hyperparameters are estimated by maximum likelihood using the Broyden-Fletcher-Goldfarh-Shanno (BFGS) optimization algorithm. The additional uncertainty of using maximum likelihood estimates for the hyperparameters in the Kalman filter is ignored in the standard errors of the filtered state variables. Since the observed time series contains 185 monthly periods, this additional uncertainty can be ignored. See also \citet{bollineni2017} for details. Both the simulation and estimation results in Sections \ref{Simulations} and \ref{empirics} are obtained using the statistical software R.

Assuming normality of the innovations is common in state space models because the hyperparameters of the model are estimated by maximizing a Gaussian log-likelihood which is evaluated by the Kalman filter. Moreover, under normality, the Kalman filter yields the minimum variance unbiased estimator of the state variables. Nonetheless, as long as the state space model is linear, if the true distribution of the error terms is non-Gaussian, then the Kalman filter still provides the minimum variance \textit{linear} unbiased estimator of the state variables \citep[Chapter~4]{durbinkoopman2012}. In this case we can further rely on quasi maximum likelihood (QML) theory in order to perform inference based on the QML estimates of the hyperparameters. This means that the hyperparameters can still be consistently estimated by maximizing the Gaussian log-likelihood (or in general, as \cite{Gourierouxetal1984} argue, a density function that belongs to the family of linear exponential distributions), but we shall use, if needed, the appropriate expression for the covariance matrix of the QML estimators, which should capture the additional uncertainty caused by the model's misspecification \citep[Chapter~13]{hamilton1994}. In Appendix \ref{simulations appendix} we conduct a Monte Carlo simulations study and find that deviations from normality are not of concern for the performance our method.


This time series model addresses and solves the mentioned problems with small sample sizes and rotation group bias. Every month a filtered estimate for the trend ($ L^{k,y}_{t}$) and the population parameter, which is defined as the filtered trend plus the filtered seasonal effect ($  \theta^{k,y}_{t} =  L^{k,y}_{t} +  S^{k,y}_{t}$), are published in month $t+1$. The time series model uses sample information from previous months in order to obtain more stable estimates. The estimates account for RGB by benchmarking the estimates for $ L^{k,y}_{t}$ and $ \theta^{k,y}_{t}$ to the level of the first wave, which makes them comparable with the outcomes obtained under the cross-sectional design before 2000.

We now introduce some further notation to distinguish between in-sample estimates and out-of-sample forecasts. In the case of in-sample estimates, $\hat \theta^{k,y}_{t|\Omega_t}$ denotes the filtered estimate of the population parameter $\theta^{k,y}_{t}$, assuming that all data for time $t$ is released and available at time $t$. We therefore condition on the information set $\Omega_t$ which does not contain any missing data at time $t$. 
In the case of out-of-sample forecasts, we condition on the data set $\Omega_t^{-}$ that is actually available in real time at time $t$. For instance, $\vy_t^k$ only gets published during moth $t+1$, and is therefore not available yet at time $t$, and not part of $\Omega_t^{-}$. 
Thus $\hat \theta^{k,y}_{t|\Omega_t^{-}}$ is the filtered forecast for $\theta^{k,y}_{t}$, based on the information that is available at time $t$. Under model (\ref{eq:univ_measurement}), which does not contain auxiliary information other than the labour force series, $\hat \theta^{k,y}_{t|\Omega_t^{-}}$ is in fact the one-step-ahead prediction $\hat \theta^{k,y}_{t|\Omega_{t-1}}$, since $\vy_t^k$ is not available yet in month $t$, but $\vy_{t-1}^k$ is; therefore, $\Omega_t^{-} = \Omega_{t-1} = \{ \vy^k_{t-1}, \vy^k_{t-2}, \ldots \}$.

\subsection{Including high-dimensional auxiliary series} \label{auxiliary}

To improve precision and timeliness of the monthly unemployment figures, we extend the labour force model by including auxiliary series of weekly/monthly Google Trends about job-search and economic uncertainty, and monthly claimant counts, in the Netherlands. Since the claimant counts for month $t$ become available in month $t+1$, it is anticipated that this auxiliary series is particularly useful to further improve the precision of the trend and population parameter estimates after finalizing the data collection for reference month $t$. The Google Trends come at a higher frequency already during the reference month $t$. It is therefore anticipated that these auxiliary series can be used to make first provisional estimates for the trend and the population parameter of the LFS during month $t$, when the sample estimates $\vy^k_t$ are not available, but the Google Trends become available on weekly basis.

Weekly and monthly Google Trends are throughout the paper denoted by $\vx^{GT}_t$ and $\vx^{k,GT}_t$, respectively. We denote the dimension of the vector $\vx^{GT}_t$ by $n$, which can be large. In addition, we can expect the Google Trends to be very noisy, such that the signal about unemployment contained in them is weak. 
We therefore need to address the high-dimensionality of these auxiliary series, in order to make the dimension of our state space model manageable for estimation, and extract the relevant information from these series. 
For this purpose we employ a factor model which achieves both by retaining the information of these time series in a few common factors.

Moreover, when dealing with mixed frequency variables and with publication delays, we can encounter ``jagged edge" datasets, which have missing values at the end of the sample period. The Kalman filter computes a prediction for the unobserved components in presence of missing observations for the respective observable variables.

The two-step estimator by \cite{Doz2011} combines factor models with the Kalman filter and hence addresses both of these issues. In the remainder of this section we explain how this estimator can be employed to nowcast the lower-frequency unobserved components of the labour force model using information from higher-frequency or real-time auxiliary series.




We consider the following state space representation of the dynamic factor model for the Google Trends, with respective measurement and transition equations, as we would like to link it to the state space model used to estimate the unemployment \eqref{eq:univ_measurement}:
\begin{equation} \label{eq:factor1}
\begin{aligned}
\vx^{GT}_t & = \mLambda \vf_t + \vvarepsilon_t, \quad \vvarepsilon_t \sim N (\vzeros, \mPsi) \\
\vf_t & = \vf_{t-1} + \vu_t, \quad \vu_t \sim N(\vzeros, \mI_r),
\end{aligned}
\end{equation}
where $\vx^{GT}_t$ is a $n\times1$ vector of observed series, $\vf_t$ is a $r \times 1$ vector of latent factors with $r \ll n$, $\mLambda$ is a $n \times r$ matrix of factor loadings, $\vvarepsilon_t$ is the $n \times 1$ vector of idiosyncratic components and $\mPsi$ its $n \times n$ covariance matrix; $\vu_t$ is the $r\times1$ vector of factors' innovations and $\mI_r$ is a $r\times r$ identity matrix (which follows from the identification conditions used in principal component analysis since the factors are only identified up to rotation). Notice that the dynamic equation for $\vf_t$ implies that we are making the assumption that $\vx^{GT}_t$ is $I(1)$ of dimension $n$, and $\vf_t$ is $I(1)$ of dimension $r$. Later in this section the need of this assumption will become clearer; the intuition behind it is that the factors and the change in unemployment, $R^{k,y}_t$, must be of the same order of integration. 

Among others, \cite{Bai2004} proves the consistency of the estimator of $I(1)$ factors by principal component analysis (PCA), under the assumptions of limited time and cross-sectional dependence and stationarity of the idiosyncratic components, $\vvarepsilon_t$, and non-trivial contributions of the factors to the variance of $\vx_t$.\footnote{For the exact formulation we refer to Assumptions A-D in \citet{Bai2004}.} We assume no cointegrating relationships among the factors. We further assume normality of the innovations for the same reasons outlined in Section \ref{DLFS}.


The consistency of the two-step estimator has been originally proven in the stationary framework by \cite{Doz2011}, and extended to the nonstationary case by \cite{Barigozzi2017}.

In the first step, the factors ($\vf_t$), the factor loadings ($\mLambda$), and the covariance matrix of the idiosyncratic components ($\mPsi$) in model \eqref{eq:factor1} are estimated by PCA as in \cite{Bai2004}. 
The matrices $\mLambda$ and $\mPsi$ are then replaced, in model \eqref{eq:factor1}, by their estimates $\hat{\mLambda}$ and $\hat{\mPsi}=\diag\left(\hat{\psi}_{11}, \dots , \hat{\psi}_{nn} \right)$ obtained in this first step. These estimates are kept as fixed in the second step, because their high-dimensionality and associated curse of dimensionality complicates re-estimation by maximum likelihood. Moreover, restricting the covariance matrix of the idiosyncratic components $\mPsi$ as being diagonal is standard in the literature\footnote{The specification of the dynamic factor factor model with spherical idiosyncratic components is often called ``approximate" dynamic factor model. \cite{Doz2011} and \cite{Barigozzi2017} mention that misspecifications of this model arising from time or cross-sectional dependence of the idiosyncratic components, do not affect the consistency of the two-step estimator of the unobserved common factors, if $n$ is large.} 


In order to make use of the auxiliary series to nowcast the unemployment, we stack together the measurement equations for $\vy^k_t$ and $\vx^{k,GT}_t$, respectively \eqref{eq:univ_measurement} and the first equation of \eqref{eq:factor1} with $\mLambda$ and $\mPsi$ replaced, respectively, by $\hat{\mLambda}$ and $\hat{\mPsi}$, and express them at the lowest frequency (in our case the monthly observation's frequency of $\vy^k_t$).
The transition equations for the RGB and survey error component in combination with the rotation scheme applied in the Dutch LFS hamper a formulation of the model on the high frequency. 
This means that $\vx^{GT}_t$ needs to be first temporally aggregated from the high to the low frequency (either before or after the first step which estimates $\mLambda$ and $\mPsi$). Since $\vx^{GT}_t$ are the $I(1)$ weekly Google Trends, which are flow variables as they measure the number of queries made during each week, they are aggregated according to the following rule \citep{banbura2013}:
\begin{equation} \label{eq:aggregation}
\vx^{k,GT}_{j,t} = \sum_{i=1}^{j} \vx^{GT}_{t-k+i}, \quad j = 1, \ldots, k, \quad t = k, 2k, \ldots \quad .
\end{equation}
The aggregated $\vx^{k,GT}_{j,t}$ are then rescaled in order to be bounded again between 0 and 100. The subscript $j$ allows for real-time updating of the aggregated Google Trends in week $j$ when new data become available. As such, this index indicates that we aggregate weeks 1 up to $j$. When $j=k$ we are at the end of the month, and we simply write $\vx^{k,GT}_{t}$ to indicate the end-of-month aggregate value.

In order to get the final model, we also include a measurement equation for the univariate auxiliary series of the claimant counts, assuming that its state vector, $\theta^{k,CC}_t$, has the same composition of our population parameter $\theta^{k,y}_t$ (i.e., composed of a smooth trend and a seasonal component):
\begin{equation} \label{eq:measurement_mixed}
 \left( \begin{array}{c} \vy^k_{t} \\ x^{k,CC}_{t} \\ \vx^{k,GT}_{t} \end{array} \right) = \left( \begin{array}{c} \vones_5  \theta^{k,y}_{t} \\ \theta^{k,CC}_{t} \\ \hat{\mLambda} \vf^k_{t} \end{array} \right)  + \left( \begin{array}{c}
\vlambda^k_t \\ 0 \\ \vzeros \end{array} \right) + \left( \begin{array}{c} \ve^k_{t} \\ \varepsilon^{k,CC}_{t} \\ \vvarepsilon^{k,GT}_{t} \end{array} \right), \quad
\left( \begin{array}{c} \varepsilon^{k,CC}_t \\ \vvarepsilon^{k,GT}_{t} \end{array} \right)
 \sim N \left( \vzeros, \left[ \begin{array}{cc} \sigma^2_{\varepsilon, CC} & \vzeros \\ \vzeros & \hat{\mPsi} \end{array} \right] \right),
\end{equation}
\begin{equation}
 \left( \begin{array}{c}\theta^{k,y}_{t} \\ \theta^{k,CC}_{t} \end{array} \right)  = \left( \begin{array}{c} L^{k,y}_{t} \\ L^{k,CC}_{t} \end{array} \right) +  \left( \begin{array}{c} S^{k,y}_{t} \\ S^{k,CC}_{t} \end{array} \right), 
 \end{equation}
 \begin{equation}
 \left( \begin{array}{c} L^{k,y}_{t} \\ L^{k,CC}_{t} \end{array} \right)  = \left( \begin{array}{c} L^{k,y}_{t-1} \\ L^{k,CC}_{t-1} \end{array} \right) + \left( \begin{array}{c} R^{k,y}_{t-1} \\ R^{k,CC}_{t-1} \end{array} \right),
\end{equation}
\begin{equation} \label{eq:slopes}
\left( \begin{array}{c} 
R^{k,y}_{t} \\ R^{k,CC}_{t} \\ \vf^k_{t}
\end{array}\right) = \left( \begin{array}{c} 
R^{k,y}_{t-1} \\ R^{k,CC}_{t-1} \\ \vf^k_{t-1}
\end{array}\right) +  \left( \begin{array}{c} 
\eta^{k,y}_{R,t} \\ \eta^{k,CC}_{R,t} \\ \vu^k_{t} 
\end{array}\right),
\end{equation}
\begin{equation} \label{eq:corr_factor}
\cov \left( \begin{array}{c} 
\eta^{k,y}_{R,t} \\
\eta^{k,CC}_{R,t} \\
\vu^k_{t} 
\end{array}\right) =
\left[ \begin{array}{ccccc} 
\sigma^2_{R,y}& \rho_{CC}\sigma_{R,y}\sigma_{R,CC} & \rho_{1,GT}\sigma_{R,y} & \dots & \rho_{r,GT}\sigma_{R,y} \\
\rho_{CC}\sigma_{R,y}\sigma_{R,CC} & \sigma^2_{R,CC} & 0 & \dots & 0 \\
\rho_{1,GT}\sigma_{R,y} & 0 & 1 & \dots & 0 \\
\vdots & \vdots & \vdots & \ddots & \vdots \\
\rho_{r,GT}\sigma_{R,y} & 0 & 0 & \dots & 1
\end{array}\right].
\end{equation}

%
%
%

The last equality allows the innovations of the trends' slopes, $R^{k,y}_t$ and $R^{k,CC}_t$, and of the factors of the Google Trends, to be correlated. \cite{harvey2000} show that there can be potential gains in precision, in terms of Mean Squared Error $(\MSE)$ of the Kalman filter estimators of $\theta^{k,y}_t$, $L^{k,y}_t$, and $R^{k,y}_t$, if the correlation parameters $|\rho|$s are large. Specifically, if $|\rho_{CC}| = 1$, then $\vy^k_t$ and $x^{k,CC}_t$ have a common slope. This means that $\vy^k_t$ and $x^{k,CC}_t$ are both $I(2)$, but there is a linear combination of their first differences which is stationary. Likewise, if $|\rho_{m,GT}| = 1$ then the $m^{\text{th}}$ factor of the Google Trends and the change in unemployment, $R^{k,y}_t$, are cointegrated (i.e., they have the same source of error). This is why we need the elements of the vector in \eqref{eq:slopes} to have the same order of integration, and it is via this correlation parameters that we exploit the auxiliary information. 


The second step of the estimation procedure consists of estimating the remaining hyperparameters of the whole state space model (equations \eqref{eq:measurement_mixed}-\eqref{eq:corr_factor}) by maximum likelihood, and applying the Kalman filter to re-estimate $\vf^k_t$ and to nowcast the variables of interest, $\theta^{k,y}_t$, $L^{k,y}_t$, and $R^{k,y}_t$, providing unemployment estimates in real-time before LFS data become available: $\hat{\theta}^{k,y}_{t|\Omega_t^{-}}$, $\hat{L}^{k,y}_{t|\Omega_t^{-}}$, and $\hat{R}^{k,y}_{t|\Omega_t^{-}}$ are the filtered nowcasts of, respectively, $\theta^{k,y}_t$, $L^{k,y}_t$, and $R^{k,y}_t$ based on the information set $\Omega_t^{-}$ available in month $t$. The information set in this case is $\Omega_t^{-} = \{\vx^{k,GT}_{t},  \vy^k_{t-1}, x^{k,CC}_{t-1}, \vx^{k,GT}_{t-1}, \ldots \}$. Note that, contrary to Section \ref{DLFS}, we now talk about ``nowcast" instead of ``forecast" of $\theta^{k,y}_t$ because a part of the data (the Google Trends) used in model \eqref{eq:measurement_mixed}-\eqref{eq:corr_factor} is now available in month $t$.  

Some remarks are in place.
First, although in Section \ref{DLFS} we mentioned that Statistics Netherlands publishes only $\hat{L}^{k,y}_t$ and $\hat{\theta}^{k,y}_t$ as official statistics for the unemployment, we are also interested in the estimation/nowcast accuracy of $R^{k,y}_t$ since it is the state variable of the labour force model that is directly related to the auxiliary series.

Second, note that in model \eqref{eq:factor1} we do not make use of the superscript $k$, meaning that the first step of the estimation can be performed on the high frequency (weekly in our empirical case) variables. 
Since in each week we can aggregate the weekly Google Trends to the monthly frequency, we can use the information available throughout the month to update the estimates of $\mLambda$ and $\mPsi$. If the correlations between the factors and the trend's slope of the target variable are large, this update should provide a more precise nowcast of 
$R^{k,y}_t$, $L^{k,y}_t$ and $\theta^{k,y}_t$. 

Third, we allow the factors of the Google Trends to be correlated with the change in unemployment and not with its level for two reasons: first, a smooth trend model is assumed for the population parameter, which means that the level of its trend does not have an innovation term. Second, it is reasonable to assume that people start looking for a job on the internet when they become unemployed, and hence their search behaviour should reflect the change in unemployment rather than its level.

Fourth, while our method to include auxiliary information in a state space model is based on the approach proposed by \cite{harvey2000}, the factors of the high-dimensional auxiliary series could also be included as regressors in the observation equation for the labour force. However, in such a model, the main part of the trend, $L_t^{k,y}$, will be explained by the auxiliary series in the regression component. As a result, the filtered estimates for $L_t^{k,y}$ will contain a residual trend instead of the trend of the unemployment. Since the filtered trend estimates are the most important target variables in the official monthly publications of the labour force, this approach is not further investigated in this paper.

Finally, we refer the reader to Appendices \ref{ssbiv}, \ref{sshd}, and \ref{ssbivhd} for a detailed state space representation of the labour force model when, respectively, a univariate, a high-dimensional or both type of auxiliary series are included. 
We further refer to Appendices \ref{sshd_lags} and \ref{sshd_seas} for an illustration on how to include the lags of the factors and how to model their cycle or seasonality, within our proposed high-dimensional state space model.

\section{Simulation study} \label{Simulations}


We next conduct a Monte Carlo simulations study in order to elucidate to which extent our proposed method can provide gains in the nowcast accuracy of the unobserved components of interest. For this purpose, we consider a simpler model than the one used for the labour force survey. Here $y^k_t$ is univariate following a smooth trend model, and $\vx^k_t$ represents the $(100 \times 1)$-dimensional auxiliary series with one common factor ($r = 1$). 
\begin{equation*} \begin{aligned}
 \left( \begin{array}{c} y^k_t \\ \vx^k_t \end{array} \right) &= \left( \begin{array}{c} L^{k}_t \\ \Lambda f^k_t \end{array} \right) + \left( \begin{array}{c} \varepsilon^{k,y}_t \\ \vvarepsilon^{k,x}_t \end{array} \right), \\
L^{k}_t &= L^{k}_{t-1} + R^{k}_{t-1}, \\
\left( \begin{array}{c} 
R^{k}_{t} \\
f^k_{t}
\end{array}\right) &= \left( \begin{array}{c} 
R^{k}_{t-1} \\
f^k_{t-1}
\end{array}\right) + \left( \begin{array}{c} 
\eta^{k}_{R,t} \\
u^k_t 
\end{array}\right), \quad
\left( \begin{array}{c} 
\eta^{k}_{R,t} \\
u^k_t 
\end{array}\right) \sim N \left( \vzeros , \left[ \begin{array}{cc}
1 & \rho \\
\rho & 1
\end{array} \right]\right).
\end{aligned} \end{equation*} 

We allow the slope and factor's innovations to be correlated, and we investigate the performance of the method for increasing values of the correlation parameter $\rho \in [0,0.2,0.4,0.6,0.8,0.9,0.99]$.
The auxiliary variable $\vx^k_t$ has the same frequency of $y^k_t$ and it is assumed that all $\vx^k_t$ are released at the same time without publication delays. 
The nowcast is done concurrently, i.e. in real-time based on a recursive scheme. This means that in each time point of the out-of-sample period, the hyperparameters of the model are re-estimated by maximum likelihood, extending the same used up to that period.
This is done in the third part of the sample, always assuming that $y^k_t$ is not available at time $t$, contrary to $\vx^k_t$. This implies that the available data set in period $t$ equals $\Omega_t^{-} = \{\vx^k_t, y^k_{t-1}, \vx^k_{t-1}, y^k_{t-2}, \ldots \}$. The sample size is $T = 150$ and the number of simulations is $n_{\text{sim}} = 500$.

We consider three specifications for the idiosyncratic components and the factor loadings:

\begin{enumerate}

\item Homoskedastic idiosyncratic components and dense loadings:
\begin{equation*}
\left( \begin{array}{c} \varepsilon^{k,y}_t \\ \vvarepsilon^{k,x}_t \end{array} \right) \sim N \left( \vzeros,0.5 \mI_{n+1} \right), \quad \Lambda \sim U \left(0,1 \right).
\end{equation*}

\item Homoskedastic idiosyncratic components and sparse loadings. The first half of the elements in the loadings are set equal to zero. This specification reflects the likely empirical case that some of the Google Trends are not related to the change in unemployment:
\begin{equation*}
\left( \begin{array}{c} \varepsilon^{k,y}_t \\ \vvarepsilon^{k,x}_t \end{array} \right) \sim N \left( \vzeros,0.5 \mI_{n+1} \right), \quad \Lambda = \left( \Lambda'_0, \Lambda'_1 \right)', \underset{50\times 1}{\Lambda_0} = \vzeros, \underset{50 \times 1}{\Lambda_1} \sim U \left(0,1 \right).
\end{equation*}

\item Heteroskedastic idiosyncratic components and dense loadings. The homoskedasticity assumption is here relaxed, again as not being realistic for the job search terms:
\begin{equation*}
\left( \begin{array}{c} \varepsilon^{k,y}_t \\ \vvarepsilon^{k,x}_t \end{array} \right) \sim N \left( \vzeros, \left( \begin{array}{cc} 0.5 & \vzeros' \\ \vzeros & \diag ( H ) \end{array} \right) \right), \quad H \sim U(0.5,10), \quad \Lambda \sim U \left(0,1 \right).
\end{equation*}

\end{enumerate}

Let $\valpha^k_t = \left(L^k_t, R^k_t, f^k_t \right)'$ denote the vector of state variables and $\hat \valpha^k_{t|\Omega_t^{-}}$ its estimates based on the information available at time $t$. The results from the Monte Carlo simulations are shown in Table \ref{tab:simulations}. We always report the $\MSFE$, together with its variance and bias components, of the Kalman filter estimator of $\valpha^k_t$, relative to the same measures calculated from the model that does not include the auxiliary series $\vx^k_t$. Recall that the latter comes down to making one-step-ahead predictions.
\begin{equation*}
\begin{aligned}
\MSFE (\hat{\valpha}^k_{t|\Omega_t^{-}}) &= \frac{1}{h} \sum_{t=T-h+1}^T \frac{1}{n_{\text{sim}}} \sum_{j=1}^{n_{\text{sim}}} \left( \hat{\valpha}_{jt|\Omega_t^{-}} - \valpha_{jt} \right) \left( \hat{\valpha}_{jt|\Omega_t^{-}} - \valpha_{jt} \right)',\\
\var (\hat{\valpha}^k_{t|\Omega_t^{-}}) &= \frac{1}{h} \sum_{t=T-h+1}^T \left( \frac{1}{n_{\text{sim}}} \sum_{j=1}^{n_{\text{sim}}} \left( \left( \hat{\valpha}_{jt|\Omega_t^{-}} - \valpha_{jt} \right) - \frac{1}{n_{\text{sim}}} \sum_{j=1}^{n_{\text{sim}}} \left( \hat{\valpha}_{jt|\Omega_t^{-}} - \valpha_{jt} \right) \right) \right. \\ 
&\qquad\left. \times \left( \left( \hat{\valpha}_{jt|\Omega_t^{-}} - \valpha_{jt} \right) - \frac{1}{n_{\text{sim}}} \sum_{j=1}^{n_{\text{sim}}} \left( \hat{\valpha}_{jt|\Omega_t^{-}} - \valpha_{jt} \right) \right)' \right), \\
\bias^2 (\hat{\valpha}^k_{t|\Omega_t^{-}}) &= \frac{1}{h} \sum_{t=T-h+1}^T \left( \frac{1}{n_{\text{sim}}} \sum_{j=1}^{n_{\text{sim}}} \left( \hat{\valpha}_{jt|\Omega_t^{-}} - \valpha_{jt} \right) \right) \left( \frac{1}{n_{\text{sim}}} \sum_{j=1}^{n_{\text{sim}}} \left( \hat{\valpha}_{jt|\Omega_t^{-}} - \valpha_{jt} \right) \right)',
\end{aligned}
\end{equation*}
where $h$ is the size of the of out-of-sample period. 

In every setting, both the bias and the variance of the $\MSFE$ tend to decrease with the magnitude of the correlation parameter. The improvement is more pronounced for the slope rather than the level of the trend. For the largest value of the correlation, with respect to the model which does not include auxiliary information, the gain in $\MSFE$ for the level and the slope is, respectively, of around 25\% and 75\%. Moreover, for low values of $\rho$, the $\MSFE$ does not deteriorate with respect to the benchmark model. This implies that our proposed method is robust to the inclusion of auxiliary information that does not have predictive power for the state variables of interest. In Appendix \ref{simulations appendix} we report and examine additional simulation results with non-Gaussian idiosyncratic components, and draw the same conclusions discussed above for the $\MSFE$ and the variance of the state variables' nowcasts. The bias instead worsens while deviating from Gaussianity, but it does not affect the $\MSFE$ as it only accounts for a small part of the latter measure. We therefore conclude that the performance of our method is overall robust to deviations from Gaussianity of the idiosyncratic components.

The decision to focus the simulation study on the nowcast (rather than the in-sample) performance of our method, is motivated by the fact that the added value of the Google Trends over the claimant counts is their real-time availability, which can be used to nowcast the unemployment. Nonetheless, for completeness, in the empirical application of the next section we report the results also for the in-sample performance of our method.

\begin{table}[h!]
\centering
\medskip
\begin{tabular}{lrrrrrrr} \toprule
      & \multicolumn{1}{c}{$\rho=0$} & \multicolumn{1}{c}{$\rho=0.2$} & \multicolumn{1}{c}{$\rho=0.4$} & \multicolumn{1}{c}{$\rho=0.6$} & \multicolumn{1}{c}{$\rho=0.8$} & \multicolumn{1}{c}{$\rho=0.9$} & \multicolumn{1}{c}{$\rho=0.99$}          \\ \midrule 
         &   \multicolumn{7}{c}{Homoskedastic idiosyncratic components and dense loadings} \\ \cmidrule(lr){2-8}         
  $\MSFE(\hat{L}^k_{t|\Omega_t^{-}})$        & 1.030 & 1.024 & 1.006 & 0.971 & 0.901 & 0.837 & 0.718 \\
  $\var(\hat{L}^k_{t|\Omega_t^{-}})$         & 1.031 & 1.025 & 1.007 & 0.971 & 0.901 & 0.837 & 0.718 \\
  $\bias^2(\hat{L}^k_{t|\Omega_t^{-}})$        & 0.775 & 0.767 & 0.756 & 0.733 & 0.692 & 0.659 & 0.567 \\
 $\MSFE(\hat{R}^k_{t|\Omega_t^{-}})$         & 1.044 & 1.017 & 0.941 & 0.806 & 0.588 & 0.427 & 0.198 \\
  $\var(\hat{R}^k_{t|\Omega_t^{-}})$        & 1.045 & 1.018 & 0.942 & 0.807 & 0.589 & 0.427 & 0.198 \\
    $\bias^2(\hat{R}^k_{t|\Omega_t^{-}})$      & 0.650 & 0.633 & 0.583 & 0.492 & 0.350 & 0.252 & 0.122 \\ \midrule
   &   \multicolumn{7}{c}{Homoskedastic idiosyncratic components and sparse loadings} \\ \cmidrule(lr){2-8} 
$\MSFE(\hat{L}^k_{t|\Omega_t^{-}})$ & 1.031 & 1.026 & 1.011 & 0.981 & 0.920 & 0.862 & 0.744 \\
 $\var(\hat{L}^k_{t|\Omega_t^{-}})$ &   1.031 & 1.026 & 1.012 & 0.981 & 0.920 & 0.862 & 0.745 \\
 $\bias^2(\hat{L}^k_{t|\Omega_t^{-}})$ &   0.784 & 0.776 & 0.762 & 0.737 & 0.695 & 0.655 & 0.582 \\
 $\MSFE(\hat{R}^k_{t|\Omega_t^{-}})$ &   1.044 & 1.019 & 0.946 & 0.817 & 0.605 & 0.446 & 0.208 \\
 $\var(\hat{R}^k_{t|\Omega_t^{-}})$ &   1.045 & 1.020 & 0.947 & 0.817 & 0.606 & 0.446 & 0.209 \\
 $\bias^2(\hat{R}^k_{t|\Omega_t^{-}})$ &  0.656 & 0.639 & 0.586 & 0.492 & 0.347 & 0.243 & 0.104 \\  \midrule
    &   \multicolumn{7}{c}{Heteroskedastic idiosyncratic components and dense loadings} \\ \cmidrule(lr){2-8}
$\MSFE(\hat{L}^k_{t|\Omega_t^{-}})$          &  1.036 & 1.032 & 1.019 & 0.994 & 0.945 & 0.901 & 0.823 \\
$\var(\hat{L}^k_{t|\Omega_t^{-}})$          &  1.037 & 1.032 & 1.020 & 0.995 & 0.946 & 0.902 & 0.823 \\
$\bias^2(\hat{L}^k_{t|\Omega_t^{-}})$          &  0.707 & 0.645 & 0.579 & 0.521 & 0.484 & 0.483 & 0.543 \\
$\MSFE(\hat{R}^k_{t|\Omega_t^{-}})$          & 1.049 & 1.027 & 0.960 & 0.840 & 0.644 & 0.499 & 0.299 \\
$\var(\hat{R}^k_{t|\Omega_t^{-}})$          & 1.049 & 1.028 & 0.961 & 0.841 & 0.645 & 0.500 & 0.299 \\
$\bias^2(\hat{R}^k_{t|\Omega_t^{-}})$          &  0.805 & 0.697 & 0.556 & 0.397 & 0.230 & 0.161 & 0.237 \\
\bottomrule
\end{tabular}
\caption{Simulation results from the three settings described in Section \ref{Simulations}. The values are reported relative to the respective measures calculated from the model that does not include the auxiliary series; values $<1$ are in favour of our method.  $n_{\text{sim}}=500$.}
\label{tab:simulations}
\end{table}

\section{Application to Dutch unemployment nowcasting} \label{empirics}

In this section we present and discuss the results of the empirical application of our method to nowcasting the Dutch unemployment using the auxiliary series of claimant counts and Google Trends related to job-search and economic uncertainty.

As explained in Section \ref{auxiliary}, the Google series used in the model must be $I(1)$. We therefore test for nonstationarity in the Google Trends with the \cite{Elliott1996} augmented Dickey-Fuller (ADF) test, including a constant and a linear trend. We control for the false discovery rate as in \cite{Moon2012}, who employ a moving block bootstrap approach that accounts for time and cross-sectional dependence among the units in the panel.

Before proceeding with the estimation of the model by only including the Google Trends that resulted as being $I(1)$ from the multiple hypotheses testing, we carry out an additional selection of the $I(1)$ Google Trends by ``targeting" them as explained and motivated below.

\cite{BAING2008target} point out that having more data to extract factors from is not always better. In particular, if series are added that have loadings of zero and are thus not influenced by the factors, these will make the estimation of factors and loadings by PCA deteriorate, as PCA assigns a non-zero weight to each series in calculating the estimated factor as a weighted average. \cite{BAING2008target} recommend a simple strategy to filter out irrelevant series (in our case Google search terms) and improve the estimation of the factors, which they call ``targeting the predictors". In this case an initial regression of the series of interest is performed on the high-dimensional input series to determine which series are (ir)relevant. The series that are found to be irrelevant are discarded and only the ones that are found to be relevant are kept to estimate the factors and loadings from. In particular, they recommend the use of the elastic net \citep{hastie2005}, which is a penalized regression technique that performs estimation and variable selection at the same time by setting the coefficients of the irrelevant variables to 0 exactly. After performing the elastic net estimation, only the variables with non-zero coefficients are then kept. As we do not observe our series of interest directly, we need to adapt their procedure to our setting. To do so we approximate the unobserved unemployment by its estimation from the labour force model without auxiliary series. Specifically, we regress the differenced estimated change in unemployment from the labour force model without auxiliary series, $\Delta \hat{R}^{k,y}_t$, on the differenced $I(1)$ Google Trends using the elastic net penalized regression method, which solves the following minimization problem:
\begin{equation*}
\min_{\vbeta} \left[ \frac{1}{2T} \sum_{t=1}^{T}\left(\Delta \hat{R}^{k,y}_t - \vbeta^\prime \Delta \vx^{k,GT}_t \right)^2 + \lambda P_\alpha \left(\vbeta \right) \right],
\end{equation*}
where 
\begin{equation*}
P_\alpha \left(\vbeta \right) = \left ( 1-\alpha \right) \frac{1}{2} || \vbeta ||^2_2 + \alpha || \vbeta ||_1.
\end{equation*}
The tuning parameters $\lambda$ and $\alpha$ are selected from a two-dimensional grid in order to minimize the \cite{Schwarz1978} Bayesian information criterion (BIC). 
Notice that performing the penalized regression on the differenced (and therefore stationary) data, also allows us to avoid the inclusion in the model of Google Trends that have spurious relations with the change in unemployment.

We both consider estimating the final model with all Google Trends included and with only the selected Google Trends included, thereby allowing us to assess the empirical effects of targeting. The final number of nonstationary Google Trends included in the model, $n$, may differ depending on whether we use the weekly Google Trends aggregated to the monthly frequency according to equation \eqref{eq:aggregation}, or the monthly Google Trends. Whenever we apply PCA, the Google Trends are first differenced and standardized.

We further need to make sure that the stationarity assumption of the idiosyncratic components is maintained. Therefore, after having estimated the factors by PCA in model \eqref{eq:factor1}, we test which of the idiosyncratic components $\vvarepsilon_t$ are $I(1)$ with an ADF test without deterministic components, by controlling for multiple hypotheses testing as in \cite{Moon2012}. The $I(1)$ idiosyncratic components are modelled as state variables in \eqref{eq:measurement_mixed}, with the following transition equation:
\begin{equation*}
\vvarepsilon^k_t = \vvarepsilon^k_{t-1} + \vxi^k_t,
\end{equation*}

\noindent with usual normality assumptions on the $\vxi^k_t$. The covariance matrix of the idiosyncratic components $\mPsi$ is therefore estimated on the levels of the $I(0)$ idiosyncratic components and the first differences of the $I(1)$ idiosyncratic components. Appendix \ref{nsidio} provides a toy example that elucidates the estimation procedure.

Finally, we notice that although the first step of the two-step estimation procedure is meant to avoid estimating $\mPsi$ and $\mLambda$ by maximum likelihood (since they are large matrices), this pre-estimation may affect the explanatory power of the Google Trends. We here propose two different ways to obtain (possibly) more accurate estimates of these two matrices: 
\begin{itemize}
\item In Section \ref{auxiliary} we mention that the first step of the two-step estimator, which estimates $\mPsi$ and $\mLambda$ by PCA, can be carried out on the weekly Google Trends (which are therefore aggregated to the monthly frequency after the first step). Since the sample size of the high frequency data is larger, using weekly Google Trends might improve the estimation accuracy of $\mPsi$ and $\mLambda$. 
\item \cite{Doz2011} argue that from the Kalman filter estimates of the factors, it is possible to re-estimate $\mPsi$ and $\mLambda$ (by least squares), which in turn can be used to re-estimate the factors, and so on. This iterative procedure is equivalent to the Expectation–Maximization (EM) algorithm, which increases the likelihood at each step and therefore converges to the maximum likelihood solution. Notice that since the Kalman filter can (in our setting) only provide monthly estimates, the iterative estimation is done on the low-frequency Google Trends.
\end{itemize}
Later in this section we check how sensitive our empirical results are to the different estimates of $\mPsi$ and $\mLambda$. For the second type of estimation method discussed above, we only perform one additional iteration of the two-step procedure due to its computational burden.

We present empirical results for the in-sample estimates and out-of-sample forecasts. With the in-sample estimates we evaluate to which extent the auxiliary series improve the precision of the published monthly unemployment estimates after finalizing the data collection. With the out-of-sample forecasts we evaluate to which extent the auxiliary series improve the precision of provisional estimates in a nowcast procedure during the period of data collection.
We always estimate four different models: the labour force model without auxiliary series (baseline), the labour force model with auxiliary series of claimant counts (CC), of Google Trends (GT) and of both (CC \& GT). We compare the latter three models to the baseline one with the in-sample and out-of-sample exercises. The period considered for the estimation starts in January 2004 and ends in May 2019 ($T=185$ months). 
The out-of-sample nowcasts are conducted in real-time (concurrently) in the last three years of the sample based on a recursive scheme: each week or month, depending on whether we use weekly or monthly Google Trends, the model, including its hyperparameters, is re-estimated on the enlarged sample now extended by the latest observations, while assuming that the current observations for the unemployed labour force and the claimant counts are missing. Analogously, when the Google Trends are first targeted with the elastic net, the targeting is re-executed in each week or month of the out-of-sample period on the updated sample. 

We define the measure of in-sample estimation accuracy $\widehat{\MSE}(\hat{\valpha}^k_{t|\Omega_t}) = \frac{1}{T-d} \sum_{t=d+1}^{T} \hat{\mP}^k_{t|\Omega_t}$, where $\hat \valpha^k_{t|\Omega_t}$ is the vector of Kalman filter estimates of the state variables, $\hat{\mP}^k_{t|\Omega_t}$ is its estimated covariance matrix in month $t$, and $d$ is the number of state variables that are needed to estimated the labour force model without auxiliary series, and that need a diffuse initialization for their estimation ($d=17$).
The measure of nowcast accuracy, $\widehat{\MSFE}(\hat{\valpha}^k_{t|\Omega_t^{-}}) = \frac{1}{h} \sum_{t=T-h+1}^{T} \hat{\mP}^k_{t|\Omega_t^{-}}$, is the average of the nowcasted covariance matrices in the $h$ prediction months. When weekly Google Trends are used, $ \hat{\mP}^k_{t|\Omega_t^{-}} = \frac{1}{k} \sum_{j=1}^{k} \hat{\mP}^k_{j|\Omega_{j,t}^{-}}$, where $\hat{\mP}^k_{j|\Omega_{j,t}^{-}}$ is the nowcasted covariance matrix for the prediction in week $j$ of month $t$, and $\Omega_{j,t}^{-} = \{\vx^{k,GT}_{j,t}, \vy^k_{t-1}, x^{k,CC}_{t-1}, \vx^{k,GT}_{t-1}, \ldots\}$ is in this case the available information set in week $j$ of month $t$. This is because the nowcast is done recursively throughout the weeks of the out-of-sample period. We always report the relative $\widehat{\text{MS(F)E}}$ with respect to the baseline model; values lower than one are in favour of our method. We note that nowcasting under the baseline model without auxiliary series and the baseline model extended with claimant counts comes down to making one-step-ahead predictions. Expressions for $\hat \valpha^k_{t|\Omega_t}$,
$\hat \valpha^k_{t|\Omega_t^{-}}$ and their covariance matrices, $\hat{\mP}^k_{t|\Omega_t}$ and $\hat{\mP}^k_{t|\Omega_t^{-}}$, are given by the standard Kalman filter recursions, see e.g. \citet[Chapter~4]{durbinkoopman2012}.

The initial values of the hyperparameters for the maximum likelihood estimation are equal to the estimates for the labour force model obtained in \cite{VanDenBrakel2015}. 
We use a diffuse initialisation of the Kalman filter for all the state variables except for the 13 state variables that define the autocorrelation structure of the survey errors, for which we use the exact initialisation of \cite{bollineni2017}.

We use the three panel information criteria proposed by \cite{baing2002} which we indicate, as in  \cite{baing2002}, with $IC_1$, $IC_2$ and $IC_3$, in order to choose how many factors of the Google Trends to include in the model\footnote{In this paper, if for instance the information criterion $IC_1$ suggests to include 2 factors, we indicate as $IC_1=2$.}. When the Google Trends are targeted with the elastic net, the information criteria suggest to include one or two factors. In the empirical analysis we check the sensitivity of the results with respect to these two different numbers of factors included in the model.

We employ a \cite{wilks1938} likelihood ratio (LR) test to assess whether the correlation parameters are significantly different from zero, and hence adding the auxiliary information might yield a significant improvement from the baseline model. Specifically, we indicate with $\rho_{CC} = 0$, $\rho_{1,GT} = 0$ and $\rho_{2,GT} = 0$ the null hypotheses for the individual insignificance of the correlation parameter with, respectively, the claimant counts, and the first and second factor (when present) of the Google Trends. With $\vrho_{GT} = \vzeros$ and $\vrho = \vzeros$ we instead indicate the null hypotheses for the joint insignificance of, respectively, the correlations with the Google Trends' factors, and all correlation parameters. 
If the true distribution of the error terms is non-Gaussian, the LR test, based on the QML estimates, does not generally keep having, under the null hypothesis, an asymptotic $\chi^2$ distribution with degrees of freedom equal to the number of restrictions. One exception is when the covariance matrix of the error terms from a regression involving observed variables, is replaced by a consistent estimator prior to the maximization of the log-likelihood \citep{GourierouxMonfort1993}. In our case, if the idiosyncratic components of the Google Trends, $\vvarepsilon^{k,GT}_t$, are the only error terms not being normally-distributed, we may fall into this exception. The covariance matrix $\mPsi$ is indeed replaced, for the maximization of the log-likelihood, by its consistent PCA estimator obtained in the first step of the two-step estimation procedure. Nonetheless, in the setting of \cite{GourierouxMonfort1993} the regressors are observed, whereas in our case the latter are the unobserved factors. Consequently, it is not trivial to asses whether our model specification indeed falls into the above-mentioned exception. A formal proof for this is beyond the scope of this paper, but in Appendix \ref{simulations appendix} we conduct a simulation study in order to obtain the finite-sample probability density of the LR test under misspecifications of the distribution of the idiosyncratic components. We conclude that the distribution of the LR test is not affected by these misspecifications. At the end of this section we show that that there is no evidence that the error terms other than $\vvarepsilon^{k,GT}_t$, are not normally-distributed. We should therefore be able perform inference based on the usual asymptotic distribution of the LR test.

Table \ref{tab:monthly} reports the estimated hyperparameters for the four models, as well as the respective value for the maximized log-likelihood, the relative measures of in and out-of-sample performance, and the p-values from the LR tests, when the monthly Google Trends are used. 

The maximum likelihood estimates for the standard error of the seasonal components' disturbance terms tend to zero, indicating that the seasonal effects are time invariant.

Recall from equation (\ref{eq:univ_sampling_error}) that the variances of the scaled sampling errors, $\sigma^2_{\nu_{j}}$, should take values close to one. 
Their estimates are divided by (1 - $\hat{\delta}^2$) and are always slightly larger than one, which is an indication that the variance estimates of the GREG estimates, used to scale the sampling errors in equation (\ref{eq:univ_sampling_error}), somewhat underestimate the real variance of the GREG estimates. 

The correlation with the claimant counts is estimated to be above 0.9, and remains large and significant when including the Google Trends. Similar conclusions can be drawn for the correlations with the Google Trends' factors, when the Google Trends are targeted with the elastic net, and 39 of them are included in the model. When the additional targeting is not applied, and the 162 $I(1)$ Google Trends are directly included in the model, the correlation parameter with the first factor of the Google Trends is instead always small and insignificant (in this setting we do not include more than one factor). Moreover, for the same number of factors, targeting the Google Trends always yield a better performance in terms of estimation and nowcast accuracy of the state variables of interest, with respect to not targeting them. For this reason, we focus the remaining analysis of the empirical results only on the targeted Google Trends.

The best results in terms of both estimation and nowcast accuracy of all the state variables, is achieved by the CC \& GT model with one factor, yielding a gain of, respectively, around 40\% and 20\% for $\hat{R}^{k,y}_{t|\Omega_t^{-}}$, and around 20\% and 25\% for both $\hat{L}^{k,y}_{t|\Omega_t^{-}}$ and $\hat{\theta}^{k,y}_{t|\Omega_t^{-}}$, with respect to the baseline model. Note that this implies that the above-mentioned model outperforms also the model that contains only the claimant counts as auxiliary series. In general, the models with Google Trends tend to achieve a better estimation and nowcast of the change in unemployment, $R^{k,y}_t$, rather than the other two state variables, with respect to the models that include the claimant counts. 

\begin{table}[h!]
  \centering
    \resizebox{\textwidth}{!}{\begin{tabular}{lrrrrrrrrrr} \toprule
          &       &       & \multicolumn{2}{c}{$n=162, IC_1=3, IC_2=1, IC_3=10$} & \multicolumn{6}{c}{Targeted GT, $n=39, IC_1=2, IC_2=1, IC_3=2$} \\ \cmidrule(lr){4-5} \cmidrule(lr){6-10}
          &       &       & \multicolumn{2}{c}{$r=1$} & \multicolumn{3}{c}{$r=1$} & \multicolumn{2}{c}{$r=2$} &  \\ \cmidrule(lr){4-5} \cmidrule(lr){6-8} \cmidrule(lr){9-10}
          & \multicolumn{1}{c}{LF} & \multicolumn{1}{c}{CC} & \multicolumn{1}{c}{GT} & \multicolumn{1}{c}{CC \& GT} & \multicolumn{1}{c}{GT} & \multicolumn{1}{c}{CC \& GT} & \multicolumn{1}{c}{CC \& GT, all corr.} & \multicolumn{1}{c}{GT} & \multicolumn{1}{c}{CC \& GT} &  \\ \midrule
        $\hat{\sigma}_{R,y}$ & 2082.652 & 2776.030 & 1995.917 & 2704.918 & 3036.281 & 2608.394 & 3447.973 & 3587.985 & 3002.947 \\
    $\hat{\sigma}_{\omega,y}$ & 0.020 & 0.020 & 0.023 & 0.078 & 0.013 & 0.011 & 0.022 & 0.054 & 0.010 \\
    $\hat{\sigma}_{\lambda}$ & 3841.035 & 3883.658 & 3592.394 & 3715.303 & 3740.097 & 3740.748 & 3115.596 & 3670.943 & 3709.361 \\
    $\hat{\sigma}_{\nu_1}$ & 1.140 & 1.151 & 1.181 & 1.146 & 1.155 & 1.142 & 1.205 & 1.155 & 1.198 \\
    $\hat{\sigma}_{\nu_2}$ & 1.291 & 1.300 & 1.270 & 1.359 & 1.276 & 1.304 & 1.378 & 1.281 & 1.263 \\
    $\hat{\sigma}_{\nu_3}$ & 1.188 & 1.181 & 1.201 & 1.211 & 1.188 & 1.196 & 1.117 & 1.224 & 1.200 \\
    $\hat{\sigma}_{\nu_4}$ & 1.240 & 1.247 & 1.241 & 1.224 & 1.241 & 1.252 & 1.356 & 1.286 & 1.243 \\
    $\hat{\sigma}_{\nu_5}$ & 1.223 & 1.228 & 1.236 & 1.260 & 1.221 & 1.239 & 1.358 & 1.254 & 1.247 \\
    $\hat{\delta}$ & 0.384 & 0.381 & 0.378 & 0.395 & 0.377 & 0.384 & 0.390 & 0.383 & 0.384 \\
     $\hat{\sigma}_{R,CC}$  &       & 3490.261 &       & 3515.222 &       & 3503.077 & 3982.583 &       & 3979.232 \\
    $\hat{\sigma}_{\omega,CC}$ &       & 0.020 &       & 0.020 &       & 0.021 & 0.016 &       & 0.020 \\
    $\hat{\sigma}_{\varepsilon,CC}$ &       & 1318.691 &       & 1310.108 &       & 1309.136 & 1181.291 &       & 1052.729 \\
    $\hat{\rho}_{CC}$ &       & 0.918 &       & 0.913 &       & 0.803 & 0.935 &       & 0.755 \\
    $\hat{\rho}_{1,GT}$ &       &       & -0.200 & -0.003 & -0.899 & -0.509 & -0.250 & -0.785 & -0.381 \\
    $\hat{\rho}_{2,GT}$ &       &       &       &       &       &       &       & -0.591 & -0.456 \\
    $\hat{\rho}_{1,CC,GT}$ &       &       &       &       &       &       & -0.093 &       &  \\ \midrule
    $\widehat{\MSE}(\hat{L}^{k,y}_{t|\Omega_t})$ &       & 0.868 & 1.003 & 0.863 & 0.919 & 0.796 & 0.895 & 0.861 & 0.849 \\
    $\widehat{\MSE}(\hat{R}^{k,y}_{t|\Omega_t})$ &       & 0.878 & 0.916 & 0.849 & 0.655 & 0.618 & 1.112 & 0.485 & 0.702 \\
    $\widehat{\MSE}(\hat{\theta}^{k,y}_{t|\Omega_t})$ &       & 0.889 & 1.009 & 0.888 & 0.941 & 0.835 & 0.916 & 0.899 & 0.881 \\ \midrule
    $\widehat{\MSFE}(\hat{L}^{k,y}_{t|\Omega_t^{-}})$ &       & 0.818 & 0.951 & 0.853 & 0.875 & 0.766 & 0.786 & 0.935 & 0.889 \\
    $\widehat{\MSFE}(\hat{R}^{k,y}_{t|\Omega_t^{-}})$ &       & 0.983 & 0.878 & 0.981 & 0.705 & 0.801 & 0.755 & 0.839 & 0.869 \\
    $\widehat{\MSFE}(\hat{\theta}^{k,y}_{t|\Omega_t^{-}})$ &       & 0.827 & 0.956 & 0.860 & 0.886 & 0.779 & 0.796 & 0.942 & 0.899 \\
    \midrule
    log-likelihood &
    -10160.378 & -11835.779 & -44726.153 & -46392.985 & -18712.139 & -20379.780 & -20384.560 & -18719.515 & -20388.495 \\ \midrule
              & \multicolumn{9}{c}{p-value from the LR test} \\ \cmidrule(lr){3-10}
        $H_0: \rho_{CC}=0$ & &  0.002     &       & 0.000 &       & 0.001 & 0.000 &       & 0.000 \\
    $H_0: \rho_{1,GT}=0$ & &       & 0.470 & 0.830 & 0.001 & 0.025 & 0.028 & 0.000 & 0.082 \\
    $H_0: \rho_{2,GT}=0$ & &       &       &       &       &       &       & 0.014 & 0.014 \\
    $H_0: \vrho_{GT}= \vzeros$ &  &      &       &       &       &       &       & 0.000 & 0.017 \\
    $H_0: \rho_{1,CC,GT}=0$ & &       &       &       &       &       & 0.470 &       &  \\
    $H_0: \vrho= \vzeros$ & &       &       & 0.001 &       & 0.000 & 0.000 &       & 0.000 \\
     \bottomrule
    \end{tabular}%
    }
    \caption{Estimation and nowcast results for the labour force model with and without auxiliary series. The auxiliary series are the claimant counts and the monthly Google Trends about job-search and economic uncertainty. The number of Google Trends and the number of their factors included in the model are denoted with $n$ and $r$, respectively. The abbreviation ``all corr." denotes that the correlation between the claimant counts and the Google Trends is also estimated. ``Targeted GT" indicates that the Google Trends have been targeted with the elastic net before including them in the model.}
  \label{tab:monthly}%
\end{table}%

Including two instead of one factor clearly increases the complexity of the model, which is reflected in smaller accuracy gains (in the CC \& GT model probably also due to the decreased magnitude of the correlation parameter with the claimant counts), especially for the nowcast of the state variables, with respect to including only one factor. Nonetheless, the correlations with both factors are individually and jointly significantly different from zero, indicating that both factors bring additional information about the Dutch unemployment.

Notice that in general all the relative measures of accuracy are below one, indicating that both the claimant counts and the Google Trends improve the estimation and nowcast accuracy of the unemployment and its change. Even when the Google Trends are not targeted and their factor is not significantly related to the unemployment, the measures are never drastically above one, meaning that our method tends to ignore auxiliary series that are not related to the target variable.

Finally, when we specified the covariance matrix \eqref{eq:corr_factor} in Section \ref{auxiliary}, we did not let the claimant counts and the Google Trends be correlated because our goal is to improve the estimation/nowcast accuracy of the unobserved components of the labour force series, not of the claimant counts nor the Google Trends. Nonetheless, if the state variables of equation \eqref{eq:slopes} are all cointegrated (i.e. the correlation parameters are all equal to one) a more efficient estimation method would be to only estimate the variance of their common source of error. We therefore estimate the CC \& GT model with one factor, when all series are correlated. We call this model ``CC \& GT all corr.". Table \ref{tab:monthly} reports the empirical results also for this model. Although the nowcast accuracy is similar to the same model without the additional correlation between the claimant counts and the Google Trends (which we indicate as $\rho_{1,CC,GT}$), the in-sample accuracy deteriorates (even with respect to the baseline model), and $\rho_{1,CC,GT}$ is not significantly different from zero. We therefore conclude that the specification of the covariance matrix \eqref{eq:corr_factor} is appropriate.

In Table \ref{tab:weekly} we report the empirical results for the GT and CC \& GT models which employ the targeted Google Trends observed at the weekly frequency, and aggregated to the monthly frequency according to equation \eqref{eq:aggregation} in order to include them in the models. In this case we still look at the sensitivity of the results with respect to the number of factors included in the model, but also with respect to the two additional methods for the estimation of $\mLambda$ and $\mPsi$ discussed at the beginning of this section. 

The measures of accuracy are again broadly lower than one, but the gains are not as large as observed for the monthly Google Trends. Including two factors improves the accuracy in the GT model, but not in the CC \& GT model, except for a more precise nowcast of $R^{k,y}_t$. The correlation parameter with the claimant counts remains large and significant. On the contrary, the correlation parameter with the first factor of the Google Trends is not significantly different from zero, and there is a weak evidence for the second factor being significantly related to the change in unemployment. For this reason we continue the analysis by considering two factors in the model.

Estimating $\mLambda$ and $\mPsi$ on the weekly Google Trends improves the measures of accuracy only for the CC \& GT model, and not for the GT model. An additional iteration of the two step estimator, in order to obtain more accurate estimates of $\mLambda$ and $\mPsi$, achieves instead better nowcasts for both the GT and the CC \& GT models (and also better in-sample estimates for the latter model), and a similar performance to the models which employ the monthly Google trends and include two factors. Notice that the values of the log-likelihood for these two models increased with respect to the same model specifications that use the original two-step estimation (without the additional iteration). The latter result, as pointed out in the explanation of the iterated estimation of $\mLambda$ and $\mPsi$ at the beginning of this section, is to be expected. Despite the above-mentioned improvements in estimation/nowcast accuracy, the correlation parameters with the Google Trends' factors are always insignificant. The aggregation of the Google Trends from the weekly to the monthly frequency yields time series that are more noisy with respect to the Google Trends that are directly observed at the monthly frequency, and detecting significant results therefore becomes harder. 

Finally, even though weekly Google Trends allow to perform the monthly nowcasts on a weekly basis, we notice that, in general, the precision of the nowcast does not monotonically improve with the number of weeks. If the high-dimensional state space model could be expressed and estimated on the highest frequency, the weekly gains in nowcast accuracy could be more evident. Nonetheless, we are limited by the transition equations for the RGB and the survey errors, to estimate the model on the monthly frequency.


\begin{table}[h!]
  \centering
    \resizebox{0.88\textwidth}{!}{
    \begin{tabular}{lrrrrrrrr}
    \toprule
          & \multicolumn{8}{c}{Targeted GT, $n=37, IC_1=1, IC_2=1, IC_3=2$} \\ \cmidrule(lr){2-9}
          & \multicolumn{2}{c}{$r=1$} & \multicolumn{6}{c}{$r=2$} \\ \cmidrule(lr){2-3} \cmidrule(lr){4-9}
          & \multicolumn{4}{c}{}          & \multicolumn{2}{c}{Weekly $\hat{\mLambda}$, $\hat{\mPsi}$} & \multicolumn{2}{c}{Iterated $\hat{\mLambda}$, $\hat{\mPsi}$} \\ \cmidrule(lr){6-7} \cmidrule(lr){8-9}
          & \multicolumn{1}{c}{GT} & \multicolumn{1}{c}{CC \& GT} & \multicolumn{1}{c}{GT} & \multicolumn{1}{c}{CC \& GT} & \multicolumn{1}{c}{GT} & \multicolumn{1}{c}{CC \& GT} & \multicolumn{1}{c}{GT} & \multicolumn{1}{c}{CC \& GT} \\ \midrule
        $\hat{\sigma}_{R,y}$ & 2020.195 & 2644.552 & 2590.937 & 3671.191 & 1995.064 & 2612.712 & 2238.557 & 2745.331 \\
    $\hat{\sigma}_{\omega,y}$ & 0.014 & 0.006 & 0.027 & 0.020 & 0.037 & 0.020 & 0.016 & 0.018 \\
    $\hat{\sigma}_{\lambda}$ & 3604.274 & 3738.299 & 3638.503 & 4281.357 & 3640.527 & 3568.508 & 3616.609 & 3635.421 \\
    $\hat{\sigma}_{\nu_1}$ & 1.146 & 1.151 & 1.142 & 1.181 & 1.161 & 1.147 & 1.155 & 1.148 \\
    $\hat{\sigma}_{\nu_2}$ & 1.295 & 1.286 & 1.292 & 1.376 & 1.294 & 1.294 & 1.278 & 1.312 \\
    $\hat{\sigma}_{\nu_3}$ & 1.203 & 1.171 & 1.208 & 1.211 & 1.167 & 1.204 & 1.207 & 1.199 \\
    $\hat{\sigma}_{\nu_4}$ & 1.253 & 1.225 & 1.248 & 1.358 & 1.247 & 1.274 & 1.252 & 1.267 \\
    $\hat{\sigma}_{\nu_5}$ & 1.240 & 1.179 & 1.234 & 1.227 & 1.244 & 1.225 & 1.231 & 1.243 \\
    $\hat{\delta}$ & 0.390 & 0.371 & 0.385 & 0.412 & 0.380 & 0.384 & 0.388 & 0.386 \\
    $\hat{\sigma}_{R,CC}$ &       & 3491.025 &       & 3635.234 &       & 3494.779 &       & 3508.248 \\
    $\hat{\sigma}_{\omega,CC}$ &       & 0.019 &       & 0.018 &       & 0.017 &       & 0.018 \\
    $\hat{\sigma}_{\varepsilon,CC}$ &       & 1342.202 &       & 1302.024 &       & 1280.971 &       & 1302.781 \\
    $\hat{\rho}_{CC}$ &       & 0.882 &       & 0.578 &       & 0.858 &       & 0.886 \\
    $\hat{\rho}_{1,GT}$ & 0.173 & -0.054 & 0.441 & -0.286 & -0.101 & -0.226 & -0.245 & 0.275 \\
    $\hat{\rho}_{2,GT}$ &       &       & 0.539 & -0.687 & 0.212 & -0.030 & 0.371 & 0.015 \\ \midrule
    $\widehat{\MSE}(\hat{L}^{k,y}_{t|\Omega_t})$ & 0.985 & 0.878 & 0.976 & 0.998 & 0.989 & 0.878 & 0.994 & 0.843 \\
    $\widehat{\MSE}(\hat{R}^{k,y}_{t|\Omega_t})$ & 0.936 & 0.872 & 0.904 & 0.996 & 0.912 & 0.836 & 0.961 & 0.799 \\
    $\widehat{\MSE}(\hat{\theta}^{k,y}_{t|\Omega_t})$ & 0.991 & 0.896 & 0.984 & 1.007 & 0.996 & 0.900 & 0.998 & 0.872 \\ \midrule
    $\widehat{\MSFE}(\hat{L}^{k,y}_{t|\Omega_t^{-}})$ & 0.990 & 0.817 & 0.909 & 0.906 & 1.008 & 0.858 & 0.914 & 0.895 \\
    week 1 & 0.988 & 0.811 & 0.928 & 0.890 & 1.005 & 0.860 & 0.909 & 0.897 \\
    week 2 & 0.989 & 0.827 & 0.894 & 0.899 & 1.015 & 0.864 & 0.910 & 0.873 \\
    week 3 & 0.993 & 0.811 & 0.901 & 0.932 & 0.993 & 0.847 & 0.895 & 0.943 \\
    week 4 & 0.995 & 0.816 & 0.911 & 0.894 & 1.011 & 0.862 & 0.948 & 0.870 \\
    week 5 & 0.969 & 0.823 & 0.920 & 0.932 & 1.032 & 0.858 & 0.897 & 0.894 \\
    $\widehat{\MSFE}(\hat{R}^{k,y}_{t|\Omega_t^{-}})$ & 0.965 & 0.982 & 0.833 & 0.843 & 0.930 & 0.840 & 0.830 & 0.819 \\
    week 1 & 0.975 & 0.981 & 0.856 & 0.860 & 0.912 & 0.843 & 0.845 & 0.839 \\
    week 2 & 0.972 & 0.991 & 0.835 & 0.832 & 0.948 & 0.862 & 0.824 & 0.812 \\
    week 3 & 0.956 & 0.954 & 0.816 & 0.832 & 0.922 & 0.831 & 0.806 & 0.821 \\
    week 4 & 0.967 & 0.987 & 0.823 & 0.844 & 0.937 & 0.817 & 0.850 & 0.811 \\
    week 5 & 0.934 & 1.021 & 0.834 & 0.852 & 0.931 & 0.867 & 0.818 & 0.794 \\
    $\widehat{\MSFE}(\hat{\theta}^{k,y}_{t|\Omega_t^{-}})$ & 0.991 & 0.825 & 0.917 & 0.937 & 0.994 & 0.873 & 0.897 & 0.902 \\
    week 1 & 0.990 & 0.820 & 0.933 & 0.943 & 1.006 & 0.876 & 0.908 & 0.894 \\
    week 2 & 0.991 & 0.835 & 0.928 & 0.963 & 0.980 & 0.873 & 0.890 & 0.886 \\
    week 3 & 0.995 & 0.820 & 0.905 & 0.933 & 1.010 & 0.860 & 0.892 & 0.959 \\
    week 4 & 0.996 & 0.825 & 0.905 & 0.908 & 1.008 & 0.884 & 0.891 & 0.882 \\
    week 5 & 0.970 & 0.830 & 0.903 & 0.944 & 0.911 & 0.867 & 0.917 & 0.871 \\ \midrule
    log-likelihood & -17954.398 & -19621.000 & -17767.456 & -19438.409 & -18651.434 & -20318.457 & -17745.335 & -19413.916 \\ \midrule
              & \multicolumn{8}{c}{p-value from the LR test} \\  \cmidrule(lr){2-9}
        $H_0: \rho_{CC}=0$ &       & 0.001 &       & 0.000 &       & 0.001 &       & 0.000 \\
    $H_0: \rho_{1,GT}=0$ & 0.813 & 0.514 & 0.689 & 1.000 & 0.555 & 1.000 & 0.685 & 1.000 \\
    $H_0: \rho_{2,GT}=0$ &       &       & 0.133 & 0.070 & 0.604 & 1.000 & 0.221 & 1.000 \\
    $H_0: \vrho_{GT}= \vzeros$ &       &       & 0.247 & 0.062 & 0.759 & 1.000 & 0.429 & 1.000 \\
    $H_0: \vrho= \vzeros$ &       & 0.001 &       & 0.001 &       & 0.004 &       & 0.002 \\
     \bottomrule
    \end{tabular}%
    }
    \caption{Estimation and nowcast results for the labour force model with auxiliary series of claimant counts and weekly Google Trends about job-search and economic uncertainty (aggregated to the monthly frequency according to equation \eqref{eq:aggregation}). The number of Google Trends and the number of their factors included in the model are denoted with $n$ and $r$, respectively. ``Weekly $\hat{\mLambda}$, $\hat{\mPsi}$" denotes that the latter estimates are obtained using the weekly Google Trends. ``Iterated $\hat{\mLambda}$, $\hat{\mPsi}$" means that the latter estimates are obtained from an additional iteration of the two-step estimator. ``Targeted GT" indicates that the Google Trends have been targeted with the elastic net before including them in the model.}
  \label{tab:weekly}%
\end{table}%

\begin{figure}[p]
\centering
\begin{tikzpicture}
\begin{axis}[ height=4.5cm, axis lines*=left, 
width=0.9\linewidth, date coordinates in=x, table/col sep=comma, date ZERO=2013-06-01, xticklabel=\year-\month, xticklabel style={rotate=30, anchor=near xticklabel},ylabel={$\hat{R}^{k,y}_{t|\Omega_t^-}$},legend style={at={(0.5,-0.5)},anchor=north,legend columns=4},axis y line*=left]
\addplot+[thick,color=brown, no markers, dashed] table[x index=0,y index=1] {nowcast_month_2factors.csv}; 
\addplot+[thick,color=red, no markers, dashed] table[x index=0,y index=4] {nowcast_month_2factors.csv}; 
\addplot+[thick,color=blue, no markers, dashed] table[x index=0,y index=7] {nowcast_month_2factors.csv}; 
\addplot+[thick,color=green, no markers, dashed] table[x index=0,y index=10] {nowcast_month_2factors.csv};
\legend{Baseline,CC,GT,CC \& GT}
\end{axis}
\end{tikzpicture}
\caption{Nowcast of $R^{k,y}_t$ with the labour force models. The results for the GT and the CC \& GT models refer to setting where the monthly Google Trends are used, and two of their factors are included in the model.} \label{fig:nowcast_R}
\end{figure}
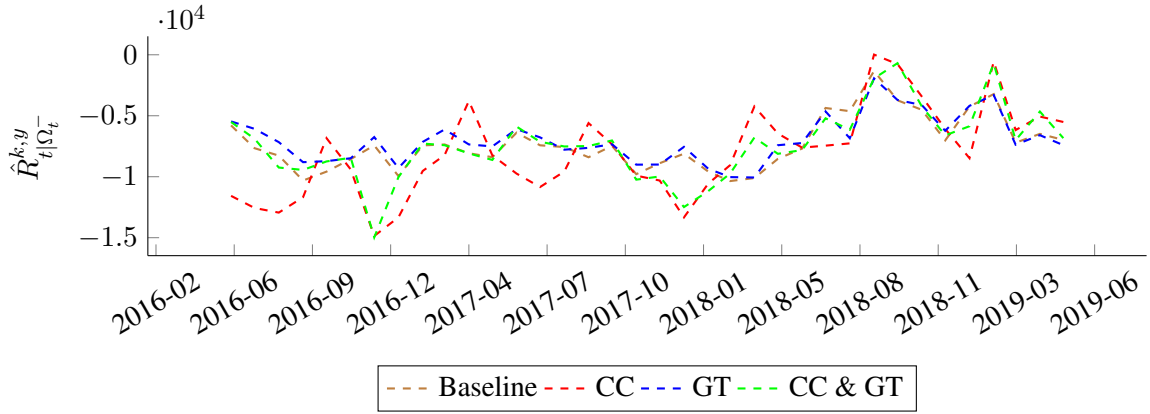

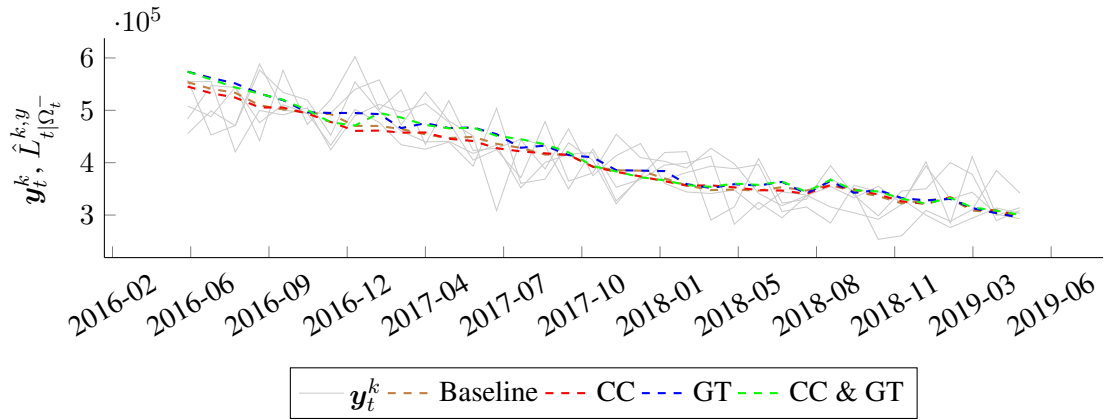
\begin{figure}[p]
\centering
\begin{tikzpicture}
\begin{axis}[ height=4.5cm, axis lines*=left,
width=0.9\linewidth, date coordinates in=x, table/col sep=comma, date ZERO=2013-06-01, xticklabel=\year-\month, xticklabel style={rotate=30, anchor=near xticklabel},ylabel={$\vy^{k}_t$, $\hat{L}^{k,y}_{t|\Omega_t^-}$},legend style={at={(0.5,-0.5)},anchor=north,legend columns=5},axis y line*=left,]
\addplot+[lgray,no markers,solid] table[x index=0,y index=13] {nowcast_month_2factors.csv}; 
\addplot+[lgray,no markers,solid] table[x index=0,y index=14] {nowcast_month_2factors.csv}; 
\addplot+[lgray,no markers,solid] table[x index=0,y index=15] {nowcast_month_2factors.csv}; 
\addplot+[lgray,no markers,solid] table[x index=0,y index=16] {nowcast_month_2factors.csv}; 
\addplot+[lgray,no markers,solid] table[x index=0,y index=17] {nowcast_month_2factors.csv}; 
\addplot+[thick,color=brown, no markers, dashed] table[x index=0,y index=2] {nowcast_month_2factors.csv}; 
\addplot+[thick,color=red,no markers, dashed] table[x index=0,y index=5] {nowcast_month_2factors.csv}; 
\addplot+[thick,color=blue, no markers, dashed] table[x index=0,y index=8] {nowcast_month_2factors.csv}; 
\addplot+[thick,color=green, no markers, dashed] table[x index=0,y index=11] {nowcast_month_2factors.csv}; 
\legend{,,,,$ \vy^{k}_t$,Baseline,CC,GT,CC \& GT}
\end{axis}
\end{tikzpicture}
\caption{Nowcast of $L^{k,y}_t$ with the labour force models, compared to the five waves of the unemployed labour force. The results for the GT and the CC \& GT models refer to the setting where the monthly Google Trends are used, and two of their factors are included in the model.} \label{fig:nowcast_L}
\end{figure}

\begin{figure}[p]
\centering
\begin{tikzpicture}
\begin{axis}[ height=4.5cm, axis lines*=left,
width=0.9\linewidth, date coordinates in=x, table/col sep=comma, date ZERO=2013-06-01, xticklabel=\year-\month, xticklabel style={rotate=30, anchor=near xticklabel},ylabel={$\vy^{k}_t$, $\hat{\theta}^{k,y}_{t|\Omega_t^-}$},legend style={at={(0.5,-0.5)},anchor=north,legend columns=5},axis y line*=left,]
\addplot+[lgray,no markers,solid] table[x index=0,y index=13] {nowcast_month_2factors.csv}; 
\addplot+[lgray,no markers,solid] table[x index=0,y index=14] {nowcast_month_2factors.csv}; 
\addplot+[lgray,no markers,solid] table[x index=0,y index=15] {nowcast_month_2factors.csv}; 
\addplot+[lgray,no markers,solid] table[x index=0,y index=16] {nowcast_month_2factors.csv}; 
\addplot+[lgray,no markers,solid] table[x index=0,y index=17] {nowcast_month_2factors.csv}; 
\addplot+[thick,color=brown, no markers, dashed] table[x index=0,y index=3] {nowcast_month_2factors.csv}; 
\addplot+[thick,color=red,no markers, dashed] table[x index=0,y index=6] {nowcast_month_2factors.csv}; 
\addplot+[thick,color=blue, no markers, dashed] table[x index=0,y index=9] {nowcast_month_2factors.csv}; 
\addplot+[thick,color=green, no markers, dashed] table[x index=0,y index=12] {nowcast_month_2factors.csv}; 
\legend{,,,,$ \vy^{k}_t$,Baseline,CC,GT,CC \& GT}
\end{axis}
\end{tikzpicture}
\caption{Nowcast of $\theta^{k,y}_t$ with the labour force models, compared to the five waves of the unemployed labour force. The results for the GT and the CC \& GT models refer to the setting where the monthly Google Trends are used, and two of their factors are included in the model.} \label{fig:nowcast_theta}
\end{figure}
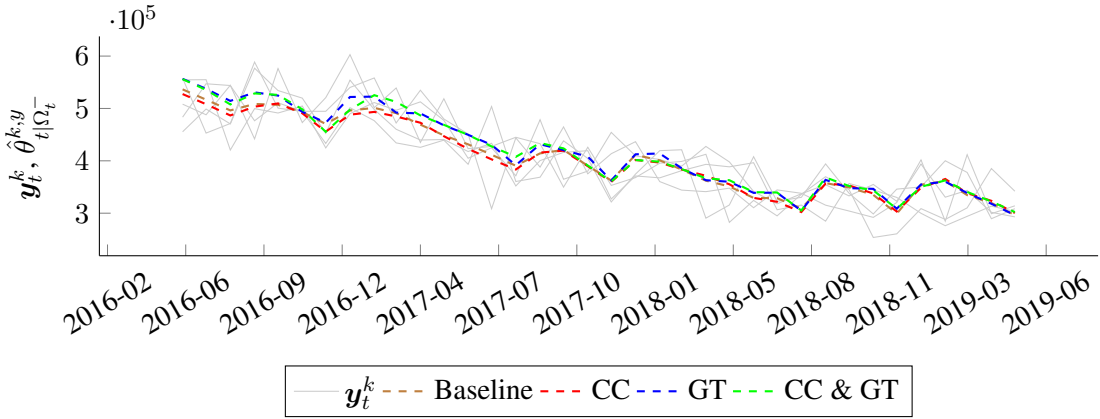

Figures \ref{fig:nowcast_R}-\ref{fig:nowcast_theta} compare the point nowcasts, respectively, of the change in unemployment, its trend, and the population parameter, obtained with the baseline, the CC, and the GT and CC \& GT models which employ monthly Google Trends and include two of their factors. From the first graph, it is evident that the models including claimant counts tend to deviate from the baseline model. The latter, on the contrary, gives similar results as those of the GT model. The point nowcasts of $L^{k,y}_t$ and $\theta^{k,y}_t$ are more similar throughout the model specifications, with a slight and positive difference between the models that include the Google Trends and the ones that do not, at the beginning of the out-of-sample period. 

Figures \ref{fig:freq_monthly} and \ref{fig:freq_agg} show the selection frequency of, respectively, the monthly and weekly Google Trends in the out-of-sample period. Some of the most selected search terms in both cases are: werklozen (unemployed people), baan zoeken (job search), curriculum vitae voorbeeld (curriculum vitae example), ww uitkering (unemployment benefits), ww aanvragen (to request unemployment benefits), resume, tijdelijk werk (temporary job), huizenmarkt zeepbel (housing market bubble). Notice that the latter term (as well as ``economische crisis" (economic crisis) or ``failliet" (bankrupt), which are also frequently selected monthly Google Trends) is of economic uncertainty nature, rather than being job-search related. A previous version of this paper only used the latter type of search terms, and did not find them to have explanatory power for the Dutch unemployment, which is now instead significantly improved by the inclusion of search terms related to economic uncertainty.

The results of the empirical analysis can be summarized as follows. Targeting the Google Trends improves the explanatory power of the latter series for the Dutch unemployment. Monthly Google Trends significantly improve the estimation and nowcast accuracy of the Dutch unemployment and its change, with both one and two factors. The largest gains are obtained when both the claimant counts and the Google Trends are included, and considering only one factor for the latter series. When two factors are considered, the gains are smaller but both factors seem to be significantly related to the change in unemployment, indicating that both of them should be included in the model in order to exploit all the information that the Google Trends give about the target variable. The sensitivity to the number of factors is somewhat similar for the weekly Google Trends, although there is a weak evidence only for their second factor to have a significant relation with the change in unemployment. The weekly Google Trends are less informative about the Dutch unemployment, yielding in general less improvements in estimation and nowcast accuracy, with respect to the monthly Google Trends. The contributions of the two types of Google Trends are comparable only when the two-step estimator is additionally re-iterated for the weekly Google Trends (in order to obtain more precise estimates of $\mLambda$ and $\mPsi$). This result suggests that iterating the two-step estimation can improve the explanatory power of the Google Trends, and that the latter series are sensitive to the estimates of $\mLambda$ and $\mPsi$. Improvements are, instead, not always present when $\mLambda$ and $\mPsi$ are estimated on the weekly data. In general, the claimant counts mainly have a positive impact on the estimation and nowcast accuracy of $\theta^{k,y}_t$ and $L^{k,y}_t$, whereas the Google Trends on $R^{k,y}_t$. The point nowcasts of the latter state variable are more sensitive to the type of auxiliary series included, with respect to the ones of $\theta^{k,y}_t$ and $L^{k,y}_t$. 

The assumptions of normality made and discussed throughout the paper can be tested on the standardized one-step ahead forecast errors \cite[Chapter 7]{durbinkoopman2012}: $\tilde{\vv}^k_t = \mB^k_t \vv^k_t$, for $t=d+1, \dots, T$ with $(\mF^k_t)^{-1}=\mB^{k'}_t \mB^k_t$, where $\mF^k_t$ is the covariance matrix of the prediction errors $\vv^k_t$ estimated with the Kalman filter. The prediction errors for the labour force are defined as $\vv^{k,y}_t = \vy^k_t - \mZ^y_t \hat{\valpha}^{k,y}_{t|\Omega_{t-1}}$, for the claimant counts as $v^{k,CC}_t = x^{k,CC}_t - \mZ^{CC} \hat{\valpha}^{k,CC}_{t|\Omega_{t-1}}$, and for the Google Trends as $\vv^{k,y}_t = \vx^{k,GT}_t - \hat{\mLambda} \hat{\vf}^{k}_{t|\Omega_{t-1}}$, for $t=d+1, \dots, T$ (the expressions for $\mZ^y_t$ and $\mZ^{CC}$ can be found in Appendix \ref{appendix_ss}). We test the assumptions on the estimated CC \& GT models when two factors of the Google Trends are included, and which employ, respectively, the monthly Google Trends, and the weekly Google Trends with the additional iteration of the two-step estimator (as they yield the best results in terms of estimation and nowcast accuracy of the state variables of interest, when two factors of the Google Trends are included).

We test the null hypothesis of univariate normality for each of the prediction error, with the \cite{ShapiroWilk1965} and \cite{BowmanShenton1975} tests, as suggested, respectively, in \citet[Chapter~5]{harvey_1990} and \citet[Chapter 2]{durbinkoopman2012}. The former test is  based on the correlation between given observations and associated normal scores, whereas the latter test is based on the measures of skewness and kurtosis. 

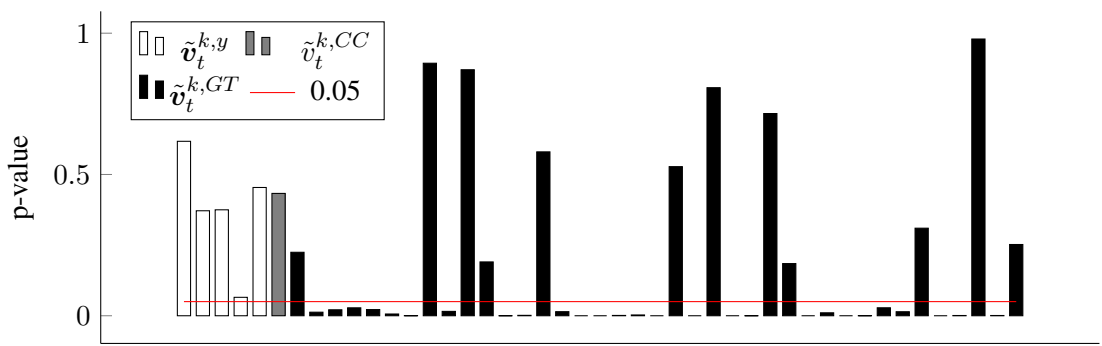
\begin{figure}[h!]
\centering
\begin{tikzpicture}
\begin{axis}[ybar=-5pt, width=0.9\linewidth, height=6cm, axis lines*=left, ylabel={p-value},
    bar width=5pt, table/col sep=comma, legend pos=north west,  legend columns=2, 
    xtick=\empty 
    ]
        \addplot[fill=white, select coords between index={0}{4}] table[y=SW, x=n]{pnorm_monthly.csv}; \addlegendentry{$\tilde{\vv}_t^{k,y}$}
        \addplot[fill=gray, select coords between index={5}{5}] table[y=SW, x=n]{pnorm_monthly.csv}; \addlegendentry{$\tilde{v}_t^{k,CC}$}
        \addplot[fill=black, select coords between index={6}{44}] table[y=SW, x=n]{pnorm_monthly.csv}; \addlegendentry{$\tilde{\vv}_t^{k,GT}$}
        \addplot[red,line legend,sharp plot] coordinates {(1,0.05) (45,0.05)}; \addlegendentry{0.05}
\end{axis}
\end{tikzpicture}
\caption{p-values from the Shapiro-Wilk test for individual normality, performed on each of the standardized prediction errors of the labour force, the claimant counts, and the Google Trends series ($\tilde{\vv}^{k}_t$). The standardized prediction errors are obtained from the CC \& GT model which employs the monthly Google Trends and include two of their factors. The red line represents the confidence level of 0.05.}
\label{fig:pvalues_norm_monthly}
\end{figure}

The p-values from the Shapiro-Wilk test are reported in Figures \ref{fig:pvalues_norm_monthly} and \ref{fig:pvalues_norm_weekly} for the two different model specifications discussed above, respectively. For both model specifications, there is no (strong) evidence against the normality assumptions for the error terms of the labour force and the claimant counts series, as their corresponding p-values are above the confidence level of 0.05. This result suggests that the model is correctly specified for these series. The test instead rejects the null hypothesis of normality for most of the idiosyncratic components of the Google Trends. The normality assumption seems therefore not appropriate for the latter series, but as discussed in Sections \ref{DLFS} and \ref{empirics}, and examined in the simulation study of Appendix \ref{simulations appendix}, this type of misspecification does not affect the consistency of the estimators of the state variables and the hyperparameters, and does not seem to influence the performance of our method, nor the distribution of the LR test which allows to perform inference on the correlation parameters\footnote{Notice that we do not control for multiple hypotheses testing in this case. If we would control for it, we would obtain less rejections of the null hypothesis of normality for the error terms of the Google Trends, but the conclusions for the error terms of the labour force and the claimant counts series would stay the same.}. The conclusions from the Bowman-Shenton test are the same and the corresponding p-values are reported in Figures \ref{fig:pvalues_norm_monthly_BS} and \ref{fig:pvalues_norm_weekly_BS}.

\begin{figure}[h!]
\centering
\begin{tikzpicture}
\begin{axis}[ybar=-5pt, width=0.9\linewidth, height=6cm, axis lines*=left, ylabel={p-value},
    bar width=5pt, table/col sep=comma, legend pos=north east,  legend columns=2, 
    xtick=\empty 
    ]
        \addplot[fill=white, select coords between index={0}{4}] table[y=SW, x=n]{pnorm_weekly.csv}; \addlegendentry{$\tilde{\vv}_t^{k,y}$}
        \addplot[fill=gray, select coords between index={5}{5}] table[y=SW, x=n]{pnorm_weekly.csv}; \addlegendentry{$\tilde{v}_t^{k,CC}$}
        \addplot[fill=black, select coords between index={6}{42}] table[y=SW, x=n]{pnorm_weekly.csv}; \addlegendentry{$\tilde{\vv}_t^{k,GT}$}
        \addplot[red,line legend,sharp plot] coordinates {(1,0.05) (43,0.05)}; \addlegendentry{0.05}
\end{axis}
\end{tikzpicture}
\caption{p-values from the Shapiro-Wilk test for individual normality, performed on each of the standardized prediction errors of the labour force, the claimant counts, and the Google Trends series ($\tilde{\vv}^{k}_t$). The standardized prediction errors are obtained from the CC \& GT model which employs the weekly Google Trends and include two of their factors, and which iterates the estimation of $\mLambda$ and $\mPsi$. The red line represents the confidence level of 0.05.}
\label{fig:pvalues_norm_weekly}
\end{figure}

\section{Conclusions} \label{conclusions}

This paper proposes a method to include a high-dimensional auxiliary series in a state space model in order to improve the estimation and nowcast of unobserved components. The method is based on a combination of PCA and Kalman filter estimation to reduce the dimensionality of the auxiliary series, originally proposed by \cite{Doz2011}, while the auxiliary information is included in the state space model as in \cite{harvey2000}. 
In this way we extend the state space model used by Statistics Netherlands to estimate the Dutch unemployment, which is based on monthly LFS data, by including the auxiliary series of claimant counts and Google Trends related to job-search and economic uncertainty. 
The strong explanatory power of the former series, in similar settings, has already been discovered in the literature (see \cite{harvey2000} and \cite{Brakel2016}). We explore to which extent a similar success can be obtained from online job-search and economic uncertainty behaviour. The advantage of Google Trends is that they are freely available at higher frequencies than the labour force survey and the claimant counts, and, contrary to the latter, they are not affected by publications delays. This feature can play a key role in the nowcast of the unemployment, as being the only real-time available information. 

A Monte Carlo simulation study shows that in a smooth trend model our proposed method can improve the $\MSFE$ of the nowcasts of the trend's level and slope up to, respectively, around 25\% and 75\%. These results are robust to misspecifications regarding the distribution of the idiosyncratic components of the auxiliary series. Therefore, our method does have the potential to improve the nowcasts of unobserved components of interest.

In the empirical application of our method to Dutch unemployment estimation and nowcasting, we find that our considered Google Trends (when first targeted with the elastic net) do in general yield gains in the estimation and nowcast accuracy (respectively up to 40\% and 25\%) of the state variables of interest, with respect to the model which does not include any auxiliary series. 
This result stresses the advantage of using the high-dimensional auxiliary series of Google Trends, despite involving a more complex model to estimate, which is especially relevant for countries that do not have any data sources related to the unemployment (such as the registry-sourced series of claimant counts), other than the labour force survey. We also find that, under certain model specifications, including both claimant counts and Google Trends outperforms the model which only includes the former auxiliary series. This result is explained by the fact that the two auxiliary series have a positive impact on the estimation/nowcast accuracy of different unobserved components which constitute the unemployment, thus yielding an overall improvement of the fit of the model. This also indicates that claimant counts and Google Trends do not bring redundant information about the Dutch unemployment.

The magnitude of the above-mentioned gains is, nonetheless, sensitive  with respect to the following aspects of the data and the model specification. First, in our empirical application we employ both monthly and weekly Google Trends. The latter need to be aggregated to the monthly frequency in order to be included in the model, but allow to perform the nowcast on a weekly basis. We find that the former are less noisy and provide in general more accurate estimates/nowcasts of the state variables of interest, with respect to the latter. The explanatory power of the monthly Google Trends for the Dutch unemployment is further corroborated by results from LR testing, which are in favour of their inclusion in the model. There is, instead, not strong and consistent evidence for this when the weekly Google Trends are employed. 

Second, PCA involves the estimation of common factors that drive the Google Trends, and in our method we relate these factors to the unobserved components that constitute the Dutch unemployment. Information criteria suggest that the Google Trends are driven by either one or two common factors. We find that including two factors yields, in general, less gains in accuracy, with respect to including one factor (due to the increased complexity of the model), but there is evidence that the second factor is related to the unemployment, and therefore it should be included in the model in order to exploit all the information that the Google Trends give about the unemployment.

Finally, our estimation method is based on a two-step procedure. In the first step, the matrix of factors' loadings and the covariance matrix of the idiosyncratic components of the Google Trends are estimated by PCA. In the second step, these matrices are replaced by their PCA estimates, in order to re-estimate the Google Trends' factors and the unobserved components of the labour force series, with the Kalman filter. Replacing these matrices by their estimates might affect the explanatory power of the Google Trends. We find that the explanatory power of the weekly Google Trends can be improved (in order to yield similar gains as the ones obtained with the monthly Google Trends), with an additional iteration of the two-step estimation procedure, which should provide more accurate estimates of the two matrices.

As already mentioned, we generally find estimation/nowcast accuracy gains from the inclusion of the Google Trends, when they are first ``targeted", by selecting the ones that are relevant for the Dutch unemployment, based on the elastic net penalized regression. If the targeting is not first applied, we do not find gains and significant relationships between the Google Trends and the Dutch unemployment. Nonetheless, in this case the results do not deteriorate with respect to the model that does not include any auxiliary series, suggesting that our method is able to ignore the inclusion of irrelevant auxiliary series, in the estimation/nowcast of unobserved components of interest. This result is corroborated in our Monte Carlo simulation study.
Hence, our proposed approach provides a framework to analyse the usefulness of ``Big Data" sources, with little risk in case the series do not appear to be useful.


One limitation of the current paper is that it does not allow for time-variation in the relation between the unobserved component of interest and the auxiliary series. For example, legislative changes may change the correlation between unemployment and administrative series such as claimant counts. Additionally, one can easily imagine the relevance of both specific search terms as well as internet search behaviour overall to change over time. While such time-variation may partly be addressed by considering shorter time periods, decreasing the already limited time dimension will have a strong detrimental effect on the quality of the estimators. Therefore, a more structural method is required that extends the current approach by building the potential for time variation into the estimation method directly, while retaining the possibility to use the full sample size. Such extensions are currently under investigation by the authors.


\renewcommand{\bibname}{References}
\bibliographystyle{apalike}
\bibliography{Paper1_Caterina_full}


\newpage

\appendix

\counterwithin{table}{section}
\counterwithin{figure}{section}


\section{State space representations} \label{appendix_ss}

For the sake of simplicity, in this appendix material the subscript $t$ (without the superscript $k$) indicates that the model is expressed at the low (monthly) frequency.


\subsection{Labour force model with univariate auxiliary series} \label{ssbiv}

Throughout this section it is assumed that the univariate auxiliary series are the claimant counts, therefore $x_t = x^{CC}_t$.

The observation equation is: 
\begin{equation*} 
\underset{6\times 1}{
\left( \begin{array}{c}
\vy_t \\
x_t
\end{array}\right)} = \mZ_t \left( \begin{array}{c}
\valpha^y_t \\
\valpha^x_t
\end{array}\right) + \left( \begin{array}{c}
\vzeros \\
\varepsilon^x_t
\end{array}\right) = \left[ \begin{array}{cc}
\mZ^y_t & \vzeros\\
\vzeros & \mZ^x 
\end{array}\right] \left( \begin{array}{c}
\valpha^y_t \\
\valpha^x_t
\end{array}\right) +\left( \begin{array}{c}
\vzeros \\
\varepsilon^x_t
\end{array}\right), \quad \left( \begin{array}{c}
\vzeros \\
\varepsilon^x_t
\end{array}\right) \sim N \left( \vzeros , 
\mH \right), \end{equation*}

$\underset{6\times 6}{\mH} = \diag \left(\vzeros', \sigma^2_{\varepsilon,x} \right)$.

The state variables for $\vy_t$ (i.e., the level, the slope, the seasonality, the RGB and the survey errors) are:
\begin{multline*}
\underset{30\times 1}{\valpha^y_t} = 
\left( \begin{matrix} L^y_t & R^y_t & S^y_{1,t} & S^{*y}_{1,t} & S^y_{2,t} & S^{*y}_{2,t} & S^y_{3,t} & S^{*y}_{3,t} & S^y_{4,t} & S^{*y}_{4,t}  \end{matrix} \right. \\
\left. \begin{matrix} S^y_{5,t} & S^{*y}_{5,t} & S^y_{6,t} & \lambda_{2,t} & \lambda_{3,t} & \lambda_{4,t} & \lambda_{5,t} & \valpha'_{E,t} \end{matrix} \right)'
\end{multline*}
\begin{equation*} \begin{aligned}
\underset{13\times 1}{\valpha_{E,t}} &= \left( \begin{array}{ccccccccccccc}
\tilde{e}_{1,t} & \tilde{e}_{2,t} & \tilde{e}_{3,t} & \tilde{e}_{4,t} & \tilde{e}_{5,t} & \tilde{e}_{1,t-2} & \tilde{e}_{2,t-2} & \tilde{e}_{3,t-2} & \tilde{e}_{4,t-2} & \tilde{e}_{1,t-1} & \tilde{e}_{2,t-1} & \tilde{e}_{3,t-1} & \tilde{e}_{4,t-1}
\end{array} \right) ',
\end{aligned} \end{equation*} 

\noindent where $E$ refers to the structure of the autocorrelated sampling errors that are modelled as state variables. 

The state variables for $x_t$ (i.e., the level, the slope and the seasonality) are:
\begin{equation*}
\underset{13\times 1}{\valpha^x_t} = \left( \begin{array}{ccccccccccccc}
L^x_t & R^x_t & S^x_{1,t} & S^{*x}_{1,t} & S^x_{2,t} & S^{*x}_{2,t} & S^x_{3,t} & S^{*x}_{3,t} & S^x_{4,t} & S^{*x}_{4,t} & S^x_{5,t} & S^{*x}_{5,t} & S^x_{6,t}
\end{array} \right) '.
\end{equation*}
\begin{equation*} \begin{aligned}
\underset{5\times 30}{\mZ^y_t} &= \left[ \begin{array}{cccccccccccccccccc}
1 & 0 & 1 & 0 & 1 & 0 & 1 & 0 & 1 & 0 & 1 & 0 & 1 & 0 & 0 & 0 & 0\\
1 & 0 & 1 & 0 & 1 & 0 & 1 & 0 & 1 & 0 & 1 & 0 & 1 & 1 & 0 & 0 & 0\\
1 & 0 & 1 & 0 & 1 & 0 & 1 & 0 & 1 & 0 & 1 & 0 & 1 & 0 & 1 & 0 & 0\\
1 & 0 & 1 & 0 & 1 & 0 & 1 & 0 & 1 & 0 & 1 & 0 & 1 & 0 & 0 & 1 & 0\\
1 & 0 & 1 & 0 & 1 & 0 & 1 & 0 & 1 & 0 & 1 & 0 & 1 & 0 & 0 & 0 & 1 
\end{array} \begin{array}{c}
\mZ^y_{E,t} \end{array} \right], \\
\underset{5\times 13}{\mZ^y_{E,t}} &= \left[ \begin{array}{ccccccccccccc}
c_{1,t} & 0 & 0 & 0 & 0 & 0 & 0 & 0 & 0 & 0 & 0 & 0 & 0 \\
0 & c_{2,t} & 0 & 0 & 0 & 0 & 0 & 0 & 0 & 0 & 0 & 0 & 0 \\
0 & 0 & c_{3,t} & 0 & 0 & 0 & 0 & 0 & 0 & 0 & 0 & 0 & 0 \\
0 & 0 & 0 & c_{4,t} & 0 & 0 & 0 & 0 & 0 & 0 & 0 & 0 & 0 \\
0 & 0 & 0 & 0 & c_{5,t} & 0 & 0 & 0 & 0 & 0 & 0 & 0 & 0 \\
\end{array} \right], \\
\underset{1\times 13}{\mZ^x} &= \left( \begin{array}{ccccccccccccc}
1 & 0 & 1 & 0 & 1 & 0 & 1 & 0 & 1 & 0 & 1 & 0 & 1
\end{array}\right).
\end{aligned} \end{equation*} 

The transition equation takes the form:
\begin{equation*}
\underset{43\times 1}{\left( \begin{array}{c}
\valpha^y_t  \\
\valpha^x_t
\end{array}\right)} = \mT \left( \begin{array}{c}
\valpha^y_{t-1} \\
\valpha^x_{t-1}
\end{array}\right) + \left( \begin{array}{c}
\veta^y_t \\
\veta^x_t
\end{array}\right) = \left[ \begin{array}{cc}
\mT^y & \vzeros\\
\vzeros & \mT^x 
\end{array}\right] \left( \begin{array}{c}
\valpha^y_{t-1} \\
\valpha^x_{t-1}
\end{array}\right) + \left( \begin{array}{c}
\veta^y_t \\
\veta^x_t
\end{array}\right).
\end{equation*}

The transition matrix for $\vy_t$ is:
\begin{equation*}
\underset{30\times 30}{\mT^y} = \blockdiag (\mT^y_{\mu}, \mT^y_{\omega},\mT^y_{\lambda}, \mT^y_{E}).
\end{equation*}

The transition matrix for the level and slope components is:
\begin{equation*}
\underset{2\times 2}{\mT^y_{\mu}} = \left[ \begin{array}{cc}
1 & 1 \\
0 & 1
\end{array}\right].
\end{equation*}

The transition matrix for the seasonal component is:
\begin{equation*}
\underset{11\times 11}{\mT^y_{\omega}} = \blockdiag (\mC_1,\mC_2,\mC_3,\mC_4,\mC_5,-1),
\end{equation*}
\begin{equation*}
\mC_j = \left[ \begin{array}{cc}
\cos(h_l) &  \sin(h_l)\\
-\sin(h_l) & \cos(h_l)
\end{array}\right], \quad h_l = \pi l / 6, \quad l=1,...,6.
\end{equation*}

The transition matrix for the RGB component is: 
\begin{equation*}
\underset{4\times 4}{\mT^y_{\lambda}} = \mI_4.
\end{equation*}

The transition matrix for the autocorrelated survey errors is: 
\begin{equation*}
\underset{13\times 13}{\mT^y_{E}} = \left[ \begin{array}{ccccccccccccc}
0 & 0 & 0 & 0 & 0 & 0 & 0 & 0 & 0 & 0 & 0 & 0 & 0 \\
0 & 0 & 0 & 0 & 0 & \delta & 0 & 0 & 0 & 0 & 0 & 0 & 0 \\
0 & 0 & 0 & 0 & 0 & 0 & \delta & 0 & 0 & 0 & 0 & 0 & 0 \\
0 & 0 & 0 & 0 & 0 & 0 & 0 & \delta & 0 & 0 & 0 & 0 & 0 \\
0 & 0 & 0 & 0 & 0 & 0 & 0 & 0 & \delta & 0 & 0 & 0 & 0 \\
0 & 0 & 0 & 0 & 0 & 0 & 0 & 0 & 0 & 1 & 0 & 0 & 0 \\
0 & 0 & 0 & 0 & 0 & 0 & 0 & 0 & 0 & 0 & 1 & 0 & 0 \\
0 & 0 & 0 & 0 & 0 & 0 & 0 & 0 & 0 & 0 & 0 & 1 & 0 \\
0 & 0 & 0 & 0 & 0 & 0 & 0 & 0 & 0 & 0 & 0 & 0 & 1 \\
1 & 0 & 0 & 0 & 0 & 0 & 0 & 0 & 0 & 0 & 0 & 0 & 0 \\
0 & 1 & 0 & 0 & 0 & 0 & 0 & 0 & 0 & 0 & 0 & 0 & 0 \\
0 & 0 & 1 & 0 & 0 & 0 & 0 & 0 & 0 & 0 & 0 & 0 & 0 \\
0 & 0 & 0 & 1 & 0 & 0 & 0 & 0 & 0 & 0 & 0 & 0 & 0 
\end{array} \right].
\end{equation*}
The transition matrix for $x_t$, $\underset{13\times 13}{\mT^x}$, is the same as $\mT^y $ without the transition matrices for the RGB component and for the survey errors.

The vector of innovations is defined as follows: 
\begin{equation*} \begin{aligned}
\underset{30\times 1}{\veta^y_t} &= \left( \begin{matrix} \eta^y_{L,t} & \eta^y_{R,t} & \eta^y_{\omega,1,t} & \eta^{*y}_{\omega,1,t} & \eta^y_{\omega,2,t} & \eta^{*y}_{\omega,2,t} & \eta^y_{\omega,3,t} & \eta^{*y}_{\omega,3,t} & \eta^y_{\omega,4,t} & \eta^{*y}_{\omega,4,t} \end{matrix} \right. \\
&\qquad \qquad \qquad \qquad \qquad \left. \begin{matrix} \eta^y_{\omega,5,t} & \eta^{*y}_{\omega,5,t} &  
\eta^y_{\omega,6,t} & \eta_{\lambda,2,t} & \eta_{\lambda,3,t} & \eta_{\lambda,4,t} & \eta_{\lambda,5,t} & \veta'^{y}_{E,t} \end{matrix} \right)',\\
\underset{13\times 1}{\veta^y_{E,t}} &= \left( \begin{array}{cccccc}
\nu_{1,t} & \nu_{2,t} & \nu_{3,t} & \nu_{4,t} & \nu_{5,t} & \vzeros'
\end{array} \right) ', \\
\underset{13\times 1}{\veta^x_t} &= \left( \begin{array}{ccccccccccccc}
\eta^x_{L,t} & \eta^x_{R,t} & \eta^x_{\omega,1,t} & \eta^{*x}_{\omega,1,t} & \eta^x_{\omega,2,t} & \eta^{*x}_{\omega,2,t} & \eta^x_{\omega,3,t} & \eta^{*x}_{\omega,3,t} & \eta^x_{\omega,4,t} & \eta^{*x}_{\omega,4,t} & \eta^x_{\omega,5,t} & \eta^{*x}_{\omega,5,t} & \eta^x_{\omega,6,t}
\end{array} \right) ', \\
\underset{43\times 1}{\veta_t} &= \left( \begin{array}{cc} \veta'^{y}_t & \veta'^{x}_t \end{array} \right) ' \sim N \left( \vzeros,\mQ \right),
\end{aligned} \end{equation*} 

\begin{equation*}
\underset{43\times 43}{\mQ} = \\ \left[ \begin{array}{ccccccccc}
\sigma^2_{L,y} & 0 & \vzeros'  & \vzeros' & \vzeros'  &  \vzeros' & 0 & 0 &  \vzeros' \\
0 & \sigma^2_{R,y} & \vzeros'  & \vzeros' & \vzeros'  &  \vzeros' & 0 & \rho \sigma_{R,y}\sigma_{R,x} &  \vzeros' \\
\vzeros & \vzeros & \mQ^y_{\omega}   & \underset{11 \times 4}{\vzeros} & \underset{11 \times 5}{\vzeros} & \underset{11 \times 8}{\vzeros} & \vzeros & \vzeros & \underset{11 \times 11}{\vzeros} \\
\vzeros & \vzeros & \underset{4 \times 11}{\vzeros}  & \mQ^y_{\lambda} & \underset{4 \times 5}{\vzeros} & \underset{4 \times 8}{\vzeros} & \vzeros & \vzeros & \underset{4 \times 11}{\vzeros}  \\
\vzeros & \vzeros & \underset{5 \times 11}{\vzeros}  & \underset{5 \times 4}{\vzeros} & \mQ^y_{\nu} &  \underset{5 \times 8}{\vzeros} & \vzeros & \vzeros &  \underset{5 \times 11}{\vzeros}  \\
\vzeros & \vzeros & \underset{8 \times 11}{\vzeros}  & \underset{8 \times 4}{\vzeros} & \underset{8 \times 5}{\vzeros}  & \underset{8 \times 8}{\vzeros} &  \vzeros & \vzeros &\underset{8 \times 11}{\vzeros} \\
0 & 0 & \vzeros'  & \vzeros' & \vzeros' &  \vzeros' & \sigma^2_{L,x}  & 0 &  \vzeros'  \\
0 & \rho\sigma_{R,y}\sigma_{R,x} & \vzeros' & \vzeros' &\vzeros'  &  \vzeros' & 0 & \sigma^2_{R,x} &\vzeros'  \\
\vzeros & \vzeros & \underset{11 \times 11}{\vzeros} & \underset{11 \times 4}{\vzeros}  & \underset{11 \times 5}{\vzeros}  & \underset{11 \times 8}{\vzeros}  &  \vzeros & \vzeros & \mQ^x_{\omega} \\
\end{array}\right],
\end{equation*}
where $\sigma^2_{L,y} = \sigma^2_{L,x} = 0$ in the Dutch labour force model, $\underset{11\times 11}{\mQ^z_{\omega}} = \sigma^2_{\omega,z} \mI_{11}$, for $z=x,y$, $\underset{4\times 4}{\mQ^y_{\lambda}} = \sigma^2_{\lambda} \mI_{4}$ and $\underset{5\times 5}{\mQ^y_{\nu}} = \diag \left( \sigma^2_{\nu_{1}}, \sigma^2_{\nu_{2}}, \sigma^2_{\nu_{3}}, \sigma^2_{\nu_{4}}, \sigma^2_{\nu_{5}} \right)$.


\subsection{Labour force model with high-dimensional auxiliary series} \label{sshd}

Throughout this section it is assumed that the high-dimensional auxiliary series are the Google Trends, therefore $\vx_t = \vx^{GT}_t$. $n$ is the number of Google Trends. It is assumed only $r=1$ factor for the Google Trends.

The observation equation is: 
\begin{equation*} \begin{aligned} 
\underset{(5+n)\times 1}{
\left( \begin{array}{c}
\vy_t \\
\vx_t
\end{array}\right)} &= \underset{(5+n)\times 31}{\mZ_t}\left( \begin{array}{c}
\valpha^y_t \\
\alpha^x_t
\end{array}\right) + \left( \begin{array}{c}
\vzeros \\
\vvarepsilon_t
\end{array}\right) = \left[ \begin{array}{cc}
\mZ^y_t & \vzeros\\
\underset{n \times 31}{\vzeros} & \underset{n\times 1}{\hat{\Lambda}}
\end{array}\right] \left( \begin{array}{c}
\valpha^y_t \\
f_t
\end{array}\right) +\left( \begin{array}{c}
\vzeros \\
\vvarepsilon_t
\end{array}\right), \quad \left( \begin{array}{c}
\vzeros \\
\vvarepsilon_t
\end{array}\right) \sim N \left( \vzeros , 
\hat{\mH} \right), \\
\underset{(5+n)\times (5+n)}{\hat{\mH}} &= \diag \left( \vzeros', \hat{\psi}_{11}, \dots , \hat{\psi}_{nn}  \right). 
\end{aligned} \end{equation*} 
$\mZ^y_t$ is the same as in Appendix \ref{ssbiv}. 

The transition equation takes the form:
\begin{equation*}
\underset{31\times 1}{\left( \begin{array}{c}
\valpha^y_t  \\
f_t
\end{array}\right)} = \underset{31\times 31}{\mT} \left( \begin{array}{c}
\valpha^y_{t-1} \\
f_{t-1}
\end{array}\right) + \left( \begin{array}{c}
\veta^y_t \\
\eta^x_t
\end{array}\right) = \left[ \begin{array}{cc}
\mT^y & \vzeros\\
\vzeros & \mT^x 
\end{array}\right] \left( \begin{array}{c}
\valpha^y_{t-1} \\
f_{t-1}
\end{array}\right) + \left( \begin{array}{c}
\veta^y_t \\
u_t
\end{array}\right).
\end{equation*}
$\mT^y$ is the same as in Appendix \ref{ssbiv}, and $\mT^x = 1$.

The vector of innovations is: 
\begin{equation*} \begin{aligned} 
\underset{31\times 1}{\veta_t} &= \left( \begin{array}{cc} \veta'^{y}_t & u_t \end{array} \right) ' \sim N \left( \vzeros,\mQ \right),\\
\underset{31\times 31}{\mQ} &=  \left[ \begin{array}{ccccccc}
\sigma^2_{L,y} & 0 & \vzeros'  & \vzeros' & \vzeros'  &  \vzeros' & 0  \\
0 & \sigma^2_{R,y} & \vzeros'  & \vzeros' & \vzeros'  &  \vzeros' &  \rho \sigma_{R,y}\sigma_u  \\
\vzeros & \vzeros & \mQ^y_{\omega}   & \underset{11 \times 4}{\vzeros} & \underset{11 \times 5}{\vzeros} & \underset{11 \times 8}{\vzeros} & \vzeros \\
\vzeros & \vzeros & \underset{4 \times 11}{\vzeros}  & \mQ^y_{\lambda} & \underset{4 \times 5}{\vzeros} & \underset{4 \times 8}{\vzeros} & \vzeros \\
0 & 0 & \underset{5 \times 11}{\vzeros}  & \underset{5 \times 4}{\vzeros} & \mQ^y_{\nu} &  \underset{5 \times 8}{\vzeros} & 0  \\
\vzeros & \vzeros & \underset{8 \times 11}{\vzeros}  & \underset{8 \times 4}{\vzeros} & \underset{8 \times 5}{\vzeros}  & \underset{8 \times 8}{\vzeros} &  \vzeros \\
0 & \rho \sigma_{R,y}\sigma_u & \vzeros'  & \vzeros' & \vzeros' &  \vzeros' & \sigma^2_u    \\
\end{array}\right],
\end{aligned} \end{equation*} 
where $\veta^{y}_t$ and the first $(30 \times 30)$ diagonal elements of $\mQ$ are the same as in Appendix \ref{ssbiv}.


 \subsubsection{Extension of the model to incorporate the lags of $f_t$} \label{sshd_lags}

Consider a regression of $\eta^y_{R,t}$ on the past values of $u_t$:
 \begin{equation*} \begin{aligned} 
 \left( \begin{array}{c}
 R^y_t \\
 f_t  
 \end{array}\right) &= \left( \begin{array}{c}
 R^y_{t-1} \\
 f_{t-1}  
  \end{array}\right) + \left( \begin{array}{c}
 \eta^y_{R,t} \\
 u_t
 \end{array}\right), \quad  u_t \sim N \left(0, \sigma_u^2 \right),\\
 \eta^y_{R,t} &= \sum_{j=1}^q \kappa_j u_{t-j} + w_t = \kappa_1 f_{t-1} +  \sum_{j=2}^q \left(\kappa_j - \kappa_{j-1} \right) f_{t-j} - \kappa_q f_{t-q-1} + w_t, \quad w_t \sim N \left(0, \sigma_w^2 \right).
 \end{aligned} \end{equation*}  

 \begin{equation*} \begin{aligned} 
 \left( \begin{array}{c}
 L^y_t \\
 R^y_t \\
 f_t \\
 f_{t-1} \\
 f_{t-2} \\
 \vdots \\
 f_{t-q}
  \end{array}\right) &=
 \left[ \begin{array}{ccccccc}
 1 & 1 & 0 & 0 & \vzeros' & 0 & 0 \\
 0 & 1 & \kappa_1 & \left(\kappa_2 - \kappa_1 \right) & \dots & \left(\kappa_q - \kappa_{q-1} \right) & - \kappa_q \\
 0 & 0 & 1 & 0 & \vzeros' & 0 & 0 \\
 0 & 0 & 1 & 0 & \vzeros' & 0 & 0 \\
 0 & 0 & 0 & 1 & \vzeros' & 0 & 0 \\
 \vzeros & \vzeros & \vzeros & \vzeros & \ddots & \vzeros & \vzeros \\
 0 & 0 & 0 & 0 & \vzeros' & 1 & 0 \\
  \end{array}\right] 
 \left( \begin{array}{c}
 L^y_{t-1} \\
 R^y_{t-1} \\
 f_{t-1} \\
 f_{t-2} \\
 f_{t-3} \\
 \vdots \\
 f_{t-q-1}
  \end{array}\right) + \left( \begin{array}{c}
 0 \\
 w_t \\
 u_t \\
 0 \\
 0 \\
 \vzeros \\
 0
  \end{array}\right), \\
 \left( \begin{array}{c}
 w_t \\
 u_t 
  \end{array}\right) &\sim N \left( \vzeros, \left[  \begin{array}{cc} \sigma_w^2 & \rho \sigma_w \sigma_u \\ \rho \sigma_w \sigma_u & \sigma_u^2  \end{array} \right] \right).
 \end{aligned} \end{equation*} 
 In the measurement equation $\mZ^x = \left[  \begin{array}{cc} \underset{n \times 1}{\hat{\Lambda}} & \underset{n \times q}{\vzeros}  \end{array} \right]$.

 \subsubsection{Extension of the model to incorporate the seasonality/cycle in $f_t$ with a (seasonal) ARIMA model} \label{sshd_seas}

Assume an ARIMA$(3,1,1)$ process for $f_t$:
 \begin{equation*}
 f_t = f_{t-1} + \phi_1 (f_{t-1} - f_{t-2}) + \phi_2 (f_{t-2} - f_{t-3}) + \phi_3 (f_{t-3} - f_{t-4}) + u_t + \gamma u_{t-1}, \quad u_t \sim N \left(0,1 \right).
 \end{equation*}

The state space representation of the above model is based on \citet[Chapter~3]{durbinkoopman2012} and illustrated below. Let $\vf_t$ be the state vector\begin{equation*}
 \vf_t = \left( \begin{array}{c} f_{t-1} \\ f_t - f_{t-1} \\ \phi_2 (f_{t-1} - f_{t-2}) \\ \phi_3 (f_{t-2} - f_{t-3}) + \gamma u_t \end{array} \right).
 \end{equation*}
 The transition equation for $\vf_t$ takes the form:
 \begin{equation*}
 \vf_t = \left[ \begin{array}{cccc} 
 1 & 1 & 0 & 0 \\
 0 & \phi_1 & 1 & 1 \\
 0 & \phi_2 & 0 & 0 \\
 0 & 0 & \frac{\phi_3}{\phi_2} & 0
 \end{array} \right] \vf_{t-1} + \left( \begin{array}{c} 0 \\ 1 \\ 0 \\ \gamma \end{array} \right) u_t.
 \end{equation*}
 Consequently, the observation equation becomes:
 \begin{equation*}
 \vx_t = \hat{\mLambda} \left( \begin{array}{cccc} 1 & 1 & 0 & 0 \end{array} \right) + \vvarepsilon_t.
 \end{equation*}
 Note that the transition equation of the full state space model is now expressed in the form:
 \begin{equation*}
 \valpha_t = \mT \valpha_{t-1} + \mR \veta_t,
 \end{equation*}
 where
 \begin{equation*}
 \underset{\dim(\valpha_t) \times \dim(\valpha_t)}{\mR} = \left[ \begin{array}{ccccc} \mI_{\dim(\valpha_t)-4} & \vzeros & \vzeros & \vzeros & \vzeros \\
 \vzeros' & 0 & 0 & 0 & 0 \\
 \vzeros' & 0 & 1 & 0 & 0 \\
 \vzeros' & 0 & 0 & 0 & 0 \\
 \vzeros' & 0 & 0 & 0 & \gamma
 \end{array} \right].
 \end{equation*}
 We here allow $u_t$ to be correlated with $\eta^y_{R,t}$.


\subsubsection{I(1) idiosyncratic components} \label{nsidio}

Consider the following toy example to have a clearer understanding of the estimation procedure when some of the idiosyncratic components are $I(1)$.
\begin{equation*}
\vx_t = \Lambda f_t + \vvarepsilon_t.
\end{equation*}

Suppose that $\vx_t$ and $\vvarepsilon_t$ are 5-dimensional vectors ($n=5$), and $f_t$ is univariate. Suppose that $\varepsilon_{1,t}$ and $\varepsilon_{3,t}$ are $I(1)$, whereas $\varepsilon_{2,t}$, $\varepsilon_{4,t}$ and $\varepsilon_{5,t}$ are $I(0)$. Then the observation equation for $\vx_t$ becomes:
\begin{equation*}
\left( \begin{array}{c}
x_{1,t} \\
x_{2,t} \\
x_{3,t} \\
x_{4,t} \\
x_{5,t}
\end{array} \right) = \left[ \begin{array}{ccc}
\Lambda_1 & 1 & 0 \\
\Lambda_2 & 0 & 0 \\
\Lambda_3 & 0 & 1 \\
\Lambda_4 & 0 & 0 \\
\Lambda_5 & 0 & 0
\end{array} \right]  \left( \begin{array}{c}
f_t \\
\varepsilon_{1,t} \\
\varepsilon_{3,t}
\end{array} \right) + \left( \begin{array}{c}
0 \\
\varepsilon_{2,t} \\
0 \\
\varepsilon_{4,t} \\
\varepsilon_{5,t}
\end{array} \right),
\end{equation*}

where $f_t$, $\varepsilon_{1,t}$  and $\varepsilon_{3,t}$ are state variables with transition equation
\begin{equation*}
\left( \begin{array}{c}
f_t \\
\varepsilon_{1,t} \\
\varepsilon_{3,t}
\end{array} \right) = \mI_3 \left( \begin{array}{c}
f_{t-1} \\
\varepsilon_{1,t-1} \\
\varepsilon_{3,t-1}
\end{array} \right) + \left( \begin{array}{c}
u_t \\
\xi_{1,t} \\
\xi_{3,t}
\end{array} \right).
\end{equation*}

\begin{equation*}
\mPsi = \cov \left( \begin{array}{ccccc}
\xi_{1,t} & \varepsilon_{2,t} & \xi_{3,t} & \varepsilon_{4,t} & \varepsilon_{5,t}
\end{array} \right)' = \cov \left( \begin{array}{ccccc}
\Delta \varepsilon_{1,t} & \varepsilon_{2,t} & \Delta \varepsilon_{3,t} & \varepsilon_{4,t} & \varepsilon_{5,t}
\end{array} \right)'.
\end{equation*}

The covariance matrix between the innovation terms in the observation equation is
\begin{equation*}
\cov \left( \begin{array}{ccccc}
0 & \varepsilon_{2,t} & 0 & \varepsilon_{4,t} & \varepsilon_{5,t}
\end{array} \right)' = \left( \begin{array}{ccccc}
0 & 0 & 0 & 0 & 0 \\
0 & \psi_{22} & 0 & 0 & 0 \\
0 & 0 & 0 & 0 & 0 \\
0 & 0 & 0 & \psi_{44} & 0 \\
0 & 0 & 0 & 0 & \psi_{55}
\end{array} \right),
\end{equation*}

and ends up in the $\mH$ matrix defined in Appendices \ref{sshd} or \ref{ssbivhd}. On the contrary, the covariance matrix between the innovations of the state variables is
\begin{equation*}
\cov \left( \begin{array}{ccc}
u_t & \xi_{1,t} & \xi_{3,t}
\end{array} \right)' = \left( \begin{array}{ccc}
1 & 0 & 0 \\
0 & \psi_{11} & 0 \\
0 & 0 & \psi_{33}
\end{array} \right),
\end{equation*} 

and ends up in the $\mQ$ matrix defined in Appendices \ref{sshd} or \ref{ssbivhd}.


\subsection{Labour force model with univariate and high-dimensional auxiliary series} \label{ssbivhd}

Throughout this section both the claimant counts and the Google Trends are included in the model as auxiliary series.

The observation equation is: 
\begin{equation*} \begin{aligned} 
\underset{(6+n)\times 1}{
\left( \begin{array}{c}
\vy_t \\
x^{CC}_t \\
\vx^{GT}_t
\end{array}\right)} &= \underset{(6+n)\times 44}{\mZ_t}\left( \begin{array}{c}
\valpha^y_t \\
\valpha^{CC}_t \\
\alpha^{GT}_t
\end{array}\right) + \left( \begin{array}{c}
\vzeros \\
\varepsilon^{CC}_t \\
\vvarepsilon_t
\end{array}\right) = \left[ \begin{array}{ccc}
\mZ^y_t & \underset{5 \times 13}{\vzeros} & \vzeros \\
\vzeros' & \underset{1\times 13}{\mZ^{CC}} & 0 \\
\underset{n \times 30}{\vzeros} & \underset{n \times 13}{\vzeros} &  \underset{n\times 1}{\hat{\Lambda}}
\end{array}\right] \left( \begin{array}{c}
\valpha^y_t \\
\valpha^{CC}_t \\
f_t
\end{array}\right) +\left( \begin{array}{c}
\vzeros \\
\varepsilon^{CC}_t \\
\vvarepsilon^{GT}_t
\end{array}\right), \\ 
\left( \begin{array}{c}
\vzeros \\
\varepsilon^{CC}_t \\
\vvarepsilon^{GT}_t
\end{array}\right) &\sim N \left( \vzeros , 
\mH \right), \qquad
\underset{(6+n)\times (6+n)}{\mH} = \diag \left(\vzeros', \sigma^2_{\varepsilon,x}, \hat{\psi}_{11}, \dots , \hat{\psi}_{nn} \right).
\end{aligned} \end{equation*} 
$\mZ^y_t$ is the same as in Appendix \ref{ssbiv}, and $\mZ^{CC}$ is the same as $\mZ^x$ in Appendix \ref{ssbiv}. 

The transition equation takes the form:
\begin{equation*}
\underset{44\times 1}{\left( \begin{array}{c}
\valpha^y_t \\
\valpha^{CC}_t \\
f_t
\end{array}\right)} = \underset{44 \times 44}{\mT} \left( \begin{array}{c}
\valpha^y_{t-1} \\
\valpha^{CC}_{t-1} \\
f_{t-1}
\end{array}\right) + \left( \begin{array}{c}
\veta^y_t \\
\veta^{CC}_t \\
\eta^{GT}_t
\end{array}\right) = \left[ \begin{array}{ccc}
\mT^y & \underset{30 \times 13}{\vzeros} & \vzeros\\
\underset{13 \times 30}{\vzeros} & \mT^{CC} & \vzeros  \\
\vzeros' & \vzeros' & 1
\end{array}\right]  \left( \begin{array}{c}
\valpha^y_{t-1} \\
\valpha^{CC}_{t-1} \\
f_{t-1}
\end{array}\right) + \left( \begin{array}{c}
\veta^y_t \\
\veta^{CC}_t \\
u_t
\end{array}\right).
\end{equation*}
$\mT^y$ is the same as in Appendix \ref{ssbiv}, and $\mT^{CC}$ and $\valpha^{CC}_t$  are, respectively, the same as $\mT^x$ and $\valpha^x_t$  in Appendix \ref{ssbiv}.

The vector of innovations is: 
\begin{small}
\begin{equation*} \begin{aligned}
\underset{44 \times 1}{\veta_t} &= \left( \begin{array}{ccc} \veta'^{y}_t & \veta'^{CC}_t & u_t \end{array} \right) ' \sim N \left( \vzeros,\mQ \right), \\
\underset{44\times 44}{\mQ} &= \left[ \begin{array}{cccccccccc}
\sigma^2_{L,y} & 0 & \vzeros'  & \vzeros' & \vzeros'  &  \vzeros' & 0 & 0 &  \vzeros' & 0\\
0 & \sigma^2_{R,y} & \vzeros'  & \vzeros' & \vzeros'  &  \vzeros' & 0 & \rho_{CC} \sigma_{R,y}\sigma_{R,CC} &  \vzeros' & \rho_{GT} \sigma_{R,y}\sigma_u \\
\vzeros & \vzeros & \mQ^y_{\omega}   & \underset{11 \times 4}{\vzeros} & \underset{11 \times 5}{\vzeros} & \underset{11 \times 8}{\vzeros} & \vzeros & \vzeros  & \underset{11 \times 11}{\vzeros} & \vzeros \\
\vzeros & \vzeros & \underset{4 \times 11}{\vzeros}  & \mQ^y_{\lambda} & \underset{4 \times 5}{\vzeros} & \underset{4 \times 8}{\vzeros} & \vzeros & \vzeros & \underset{4 \times 11}{\vzeros}  & \vzeros \\
\vzeros & \vzeros & \underset{5 \times 11}{\vzeros}  & \underset{5 \times 4}{\vzeros} & \mQ^y_{\nu} &  \underset{5 \times 8}{\vzeros} & \vzeros & \vzeros &  \underset{5 \times 11}{\vzeros}   & \vzeros \\
\vzeros & \vzeros & \underset{8 \times 11}{\vzeros}  & \underset{8 \times 4}{\vzeros} & \underset{8 \times 5}{\vzeros}  & \underset{8 \times 8}{\vzeros} &  \vzeros & \vzeros &\underset{8 \times 11}{\vzeros} & \vzeros  \\
0 & 0 & \vzeros'  & \vzeros' & \vzeros' &  \vzeros' & \sigma^2_{L,CC}  & 0 &  \vzeros' & 0 \\
0 & \rho_{CC}\sigma_{R,y}\sigma_{R,CC} & \vzeros' & \vzeros' & \vzeros' &  \vzeros' & 0 & \sigma^2_{R,CC} &\vzeros'  & 0 \\
\vzeros & \vzeros & \underset{11 \times 11}{\vzeros} & \underset{11 \times 4}{\vzeros} & \underset{11 \times 5}{\vzeros} & \underset{11 \times 8}{\vzeros} &  \vzeros & \vzeros & \mQ^{CC}_{\omega} & \vzeros \\
0 & \rho_{GT} \sigma_{R,y}\sigma_u & \vzeros' & \vzeros' &  \vzeros'  &\vzeros' &  0 & 0 & \vzeros' & \sigma^2_u \\
\end{array}\right],
\end{aligned} \end{equation*} 
\end{small}
where $\veta^{y}_t$ is the same as in Appendix \ref{ssbiv}. $\veta^{CC}_t$ and $\sigma_{R,CC}$ are respectively the same as $\veta^{x}_t$ and $\sigma_{R,x}$ in Appendix \ref{ssbiv}. The first $(43 \times 43)$ elements of $\mQ$ are the same as in  Appendix \ref{ssbiv}, whereas the last row and column are the same as in Appendix \ref{sshd}.


\newpage

\section{List of Google Trends} \label{appendix data}

\begin{table}[H]
\centering
\medskip
\resizebox{\textwidth}{!}{
\begin{tabular}{ll|ll} \toprule
Search term & Translation/explanation & Search term & Translation/explanation  \\ \midrule 
         \multicolumn{1}{l}{aanpassing} & \multicolumn{1}{l|}{adjustment} & job interview &  \\
    \multicolumn{1}{l}{aanvragen uitkering} & \multicolumn{1}{l|}{to apply for benefit} & job vacancy &  \\
    \multicolumn{1}{l}{adecco} & \multicolumn{1}{l|}{Adecco is an employment agency} & jobbird & Jobbird is a website to look for a job \\
    \multicolumn{1}{l}{advertentie plaatsen} & \multicolumn{1}{l|}{to place an advertisement} & jobbird vacatures & Jobbird vacancies \\
    \multicolumn{1}{l}{adverteren} & \multicolumn{1}{l|}{to announce} & jobnet & Jobnet is a website to look for a job \\
    \multicolumn{1}{l}{arbeidsbureau} & \multicolumn{1}{l|}{employment office} & jobs  &  \\
    \multicolumn{1}{l}{automatische incasso} & \multicolumn{1}{l|}{automatic collection of money} & jobtrack &  \\
    \multicolumn{1}{l}{baan} & \multicolumn{1}{l|}{job} & juridische vacatures & legal vanacies \\
    \multicolumn{1}{l}{baan zoeken} & \multicolumn{1}{l|}{job search} & kantonrechter & cantonal judge \\
    \multicolumn{1}{l}{banen} & \multicolumn{1}{l|}{jobs} & kantonrechtersformule & cantonal court formula (to treat e.g. severance payments) \\
    \multicolumn{1}{l}{bedrijven failliet} & \multicolumn{1}{l|}{businesses bankrupt} & maatschappelijk werk & social work \\
    \multicolumn{1}{l}{belegger.nl} & \multicolumn{1}{l|}{website about investments' information} & manpower & Manpower is an employment agency \\
    \multicolumn{1}{l}{bezuinigen} & \multicolumn{1}{l|}{to economize} & maximum dagloon & maximum daily wage \\
    \multicolumn{1}{l}{bijscholen} & \multicolumn{1}{l|}{retraining} & mijn uwv & my uwv \\
    \multicolumn{1}{l}{bijstand} & \multicolumn{1}{l|}{assistance} & modernisering & modernization \\
    \multicolumn{1}{l}{bijstandsuitkering} & \multicolumn{1}{l|}{social assistance benefit} & monsterboard & Monsterboard is a website to look for a job \\
    \multicolumn{1}{l}{collectief ontslag} & \multicolumn{1}{l|}{collective dismissal} & monsterboard vacatures & Monsterboard vacancies \\
    \multicolumn{1}{l}{creyfs} & \multicolumn{1}{l|}{Creyfs is an empolyment agency} & motivatiebrief & motivation letter \\
    \multicolumn{1}{l}{curriculum vitae} &       & motivatiebrief schrijven & to write a motivation letter \\
    \multicolumn{1}{l}{curriculum vitae template} &       & motivatiebrief voorbeeld & example of motivation letter \\
    \multicolumn{1}{l}{curriculum vitae voorbeeld} & \multicolumn{1}{l|}{curriculum vitae example} & motivation letter &  \\
    \multicolumn{1}{l}{cv} &       & nationale vacaturebank & national job bank \\
    \multicolumn{1}{l}{cv maken} & \multicolumn{1}{l|}{to make a cv} & olympia uitzendbureau & Olympia employment agency \\
    \multicolumn{1}{l}{cv maken voorbeeld} & \multicolumn{1}{l|}{to make a cv example} & omscholen & retrain \\
    \multicolumn{1}{l}{dagloon} & \multicolumn{1}{l|}{daily wage} & ondernemingsplan voorbeeld & business plan example \\
    \multicolumn{1}{l}{duur ww} & \multicolumn{1}{l|}{duration of the unemployment benefit} & ontslag & dismissal \\
    \multicolumn{1}{l}{economische crisis} & \multicolumn{1}{l|}{economic crisis} & ontslagaanvraag & dismissal application \\
    \multicolumn{1}{l}{failliet} & \multicolumn{1}{l|}{bankrupt} & ontslagprocedure & dismissal procedure \\
    \multicolumn{1}{l}{faillisementen} & \multicolumn{1}{l|}{bankruptcies} & ontslagvergoeding & severance pay \\
    \multicolumn{1}{l}{fulltime baan} & \multicolumn{1}{l|}{full-time job} & ontslagvergunning & dismissal permit \\
    \multicolumn{1}{l}{functieomschrijving} & \multicolumn{1}{l|}{job description} & open sollicitatiebrief & open application letter \\
    \multicolumn{1}{l}{geen werk} & \multicolumn{1}{l|}{no work} & partijhandel & stock trading \\
    \multicolumn{1}{l}{hoofdbedrijfschap ambachten} & \multicolumn{1}{l|}{main business crafts} & productiemedewerker & production employee \\
    \multicolumn{1}{l}{hoogte ww} & \multicolumn{1}{l|}{level of unemployment benefit} & promotiewerk & promotional work \\
    \multicolumn{1}{l}{huizenmarkt zeepbel} & \multicolumn{1}{l|}{housing market bubble} & randstad & Randstad is an employment agency \\
    \multicolumn{1}{l}{ict vacatures} & \multicolumn{1}{l|}{IT vacancies} & randstad jobs &  \\
    \multicolumn{1}{l}{ik zoek werk} & \multicolumn{1}{l|}{I am looking for a job} & randstad uitzendbureau & Randstad employment agency \\
    \multicolumn{1}{l}{indeed} & \multicolumn{1}{l|}{Indeed is a website to look for a job} & randstad vacatures & Randstad vacancies \\
    \multicolumn{1}{l}{indeed jobs} &       & receptioniste & receptionist \\
    \multicolumn{1}{l}{indeed uitzendbureau} & \multicolumn{1}{l|}{Indeed employment agency} & recht op ww & right to unemployment \\
    \multicolumn{1}{l}{indeed vacatures} & \multicolumn{1}{l|}{Indeed vacancies} & reorganisatie & reorganization \\
    \multicolumn{1}{l}{ing direct} & \multicolumn{1}{l|}{website of the ING bank} & restructuring &  \\
    \multicolumn{1}{l}{interim} & \multicolumn{1}{l|}{Interim is an employment agency} & resume &  \\
    \multicolumn{1}{l}{job} &       & resumé &  \\
    \multicolumn{1}{l}{job bird} & \multicolumn{1}{l|}{Jobbird is a website to look for a job} & resume example &  \\
    \multicolumn{1}{l}{job description} &       & resume template &  \\
          &       & salarisadministrateur & payroll administrator \\
    
\bottomrule
\end{tabular}
}
\caption{List of Google search terms and their translations/explanations (part 1).}
\label{tab:google_searches1}
\end{table}

\begin{table}[H]
\centering
\medskip
\resizebox{\textwidth}{!}{
\begin{tabular}{ll|ll} \toprule
Search term & Translation/explanation & Search term & Translation/explanation  \\ \midrule 
            schoonmaakwerk & cleaning work & vacatures limburg & jobs in Limburg (Dutch province) \\
    schuldsanering & debt restructuring & vacatures noord brabant & jobs in North Brabant (Dutch province) \\
    sociaal plan & social plan & vacatures zorg & vacancies care \\
    sollicitatie & job application & vakantiebaan & \textcolor[rgb]{ .133,  .133,  .133}{vacation job} \\
    sollicitatiebrief & cover letter & vakantiewerk & \textcolor[rgb]{ .133,  .133,  .133}{holidayjob} \\
    sollicitatiebrief schrijven & to write a cover letter & verkoopmedewerker & sales employee \\
    sollicitatiebrief voorbeeld & cover letter example & voorbeeld cv & example cv \\
    sollicitatiegesprek & job interview & voorbeeld motivatiebrief & example of motivation letter \\
    sollicitaties & job applications & vrijwilligerswerk & volunteer work \\
    solliciteren & to apply & vrijwilligerswerk buitenland & volunteering abroad \\
    solliciteren bij & apply at & werk gezocht & job search \\
    start people & Start People is an employment agency & werk in & work in \\
    start uitzendbureau & Start employment agency & werk nl & work NL \\
    tempo team & Tempo Team is an employment agency & werk.nl & website for job placement \\
    tempo-team & Tempo Team is an employment agency & werk rotterdam & work Rotterdam \\
    tempo team uitzendbureau & Tempo Team employment agency & werk utrecht & work Utrecht \\
    tempo team vacatures & Tempo Team vacancies & werk vacature & job vacancy \\
    tempoteam & Tempo Team is an employment agency & werk vacatures & job vacancies \\
    tence & Tence is an employment agency & werk vinden & to find a job \\
    tijdelijk werk & temporary job & werk zoeken & to look for a job \\
    uitkering & payment & werkbedrijf & operating company \\
    uitkering aanvragen & to claim benefits & werkeloos & unemployed \\
    uitzendbureau & employment agency & werken bij & to work at \\
    uitzendbureau amsterdam & employment agency Amsterdam & werken in & to work in \\
    uitzendbureau den haag & employment agency The Hague & werking & working \\
    uitzendbureaus & employment agencies & werkloos & unemployed \\
    uwv   & Employee Insurance Agency & werkloosheid & unemployment \\
    uwv uitkering & Employee Insurance Agency payment & werkloosheidsuitkering & unemployment benefits \\
    uwv vacatures & Employee Insurance Agency vacancies & werkloosheidswet & unemployment law \\
    uwv werkbedrijf & Employee Insurance Agency operating company & werkloze & unemployed person \\
    uwv ww & Employee Insurance Agency unemployment benefits & werklozen & unemployed people \\
    vacature & job offer & werkzoekende & job seeker \\
    vacature amsterdam & job offer Amsterdam & wet op de ondernemingsraden & \textcolor[rgb]{ .133,  .133,  .133}{Works Councils Act} \\
    vacature eindhoven & job offer Eindhoven & wholesale &  \\
    vacature secretaresse & vacancy secretary & Ww    & unemployment benefits \\
    vacaturebank & job bank & ww    & unemployment benefits \\
    vacatures & job offers & ww aanvragen & to request unemployment benefits \\
    vacatures beveiliging & job security & ww uitkering & unemployment benefit payments \\
    vacatures bouw & job construction & ww-uitkering & unemployment benefit payments \\
    vacatures brabant & jobs in Brabant (Dutch province) & ww uitkering aanvragen & claim benefits \\
    vacatures communicatie & vacancies communication & ww uitkering aanvragen uwv & claim benefits Employee Insurance Agency \\
    vacatures flevoland & jobs in Flevoland (Dutch province) & www.asnbank.nl & website of ASN bank \\
    vacatures friesland & jobs in Friesland (Dutch province) & www.uwv.nl & website of the Employee Insurance Agency \\
    vacatures horeca & vacancies hospitality & zeepbel & bubble \\
    vacatures in de zorg & vacancies in healthcare & zoek werk & search for work \\
\bottomrule
\end{tabular}
}
\caption{List of Google search terms and their translations/explanations (part 2).}
\label{tab:google_searches2}
\end{table}

From the set of search terms listed above we discard the Google Trends which have zero values for more than half of the time, before performing the empirical analysis. The final dataset is composed of 182 monthly Google Trends and 173 weekly Google Trends.


\newpage

\section{Simulation results with non-Gaussian idiosyncratic components} \label{simulations appendix}

We conduct an additional simulation study in order to assess to which extent the Gaussianity assumptions made on the innovations of the state space model influences the performance of our method. The setting of this additional study is the same as the one discussed in Section \ref{Simulations}, with the only difference that the nowcast is done in the last period of the sample for 1000 simulation runs. We consider two additional specifications that allow the idiosyncratic components to have distributions that deviate from the Gaussian one, respectively in terms of skeweness and heaviness of the tails:

\begin{enumerate}

\item Gaussian-distributed idiosyncratic components:
\begin{equation*}
\left( \begin{array}{c} \varepsilon^{k,y}_t \\ \vvarepsilon^{k,x}_t \end{array} \right) \sim N \left( \vzeros, 0.5 \mI_{n+1} \right).
\end{equation*}

Notice that this specification is the same as the first one considered in the Section \ref{Simulations}.

\item Exponentially-distributed idiosyncratic components. 
\begin{equation*}
\varepsilon^{k,y}_t  \sim N \left( 0, 0.5 \right), \quad \varepsilon^{k,x}_{i,t} \stackrel{iid}{\sim} \text{ Exp} (1), \quad \text{ for } i=1,\dots,n.
\end{equation*}

The exponential distribution is skewed with respect to the Gaussian one.

\item Student's $t$-distributed idiosyncratic components:
\begin{equation*}
\varepsilon^{k,y}_t  \sim N \left( 0, 0.5 \right), \quad \varepsilon^{k,x}_{i,t} \stackrel{iid}{\sim} t_4, \quad \text{ for } i=1,\dots,n.
\end{equation*}

The $t$ distribution with 4 degrees of freedom has heavier tails with respect to the Gaussian one.

\end{enumerate}

In all specifications $\Lambda \sim U \left(0,1 \right)$. The generated innovations according to specifications 2 and 3 above, are then standardized to make sure that their distribution is centered around 0 and their variance is equal to 0.5. This ensures that the simulations results are directly comparable with specification 1, and that any deterioration or improvement in the performance of the method can only be attributed to the non-Gaussianity of the innovations.

The additional simulation results are reported in Table \ref{tab:simulations_NG} (we report the same measures of nowcast accuracy used in the simulation study of Section \ref{Simulations}). In terms of $\MSFE$ and variance of the nowcasts of the state variables, the distribution does not seem to play a major role. For every specification these two measures improve with a similar magnitude as the correlation parameter increases. The gains are larger for the slope rather than the level of the trend. Their values, relative to the model that does not include any auxiliary series, are broadly lower than one (being around one only when the correlation parameter is small). These results are in line with the ones discussed in Section \ref{Simulations}. The squared bias, instead, seems to be much more affected by the distribution, as it worsens while deviating from Guassianity and does not improve with a larger correlation parameter. Nonetheless, 
we notice from Table \ref{tab:simulations_NG} that this deterioration of the squared bias has a minor impact on the $\MSFE$ since the latter measure is largely composed of its variance component. 

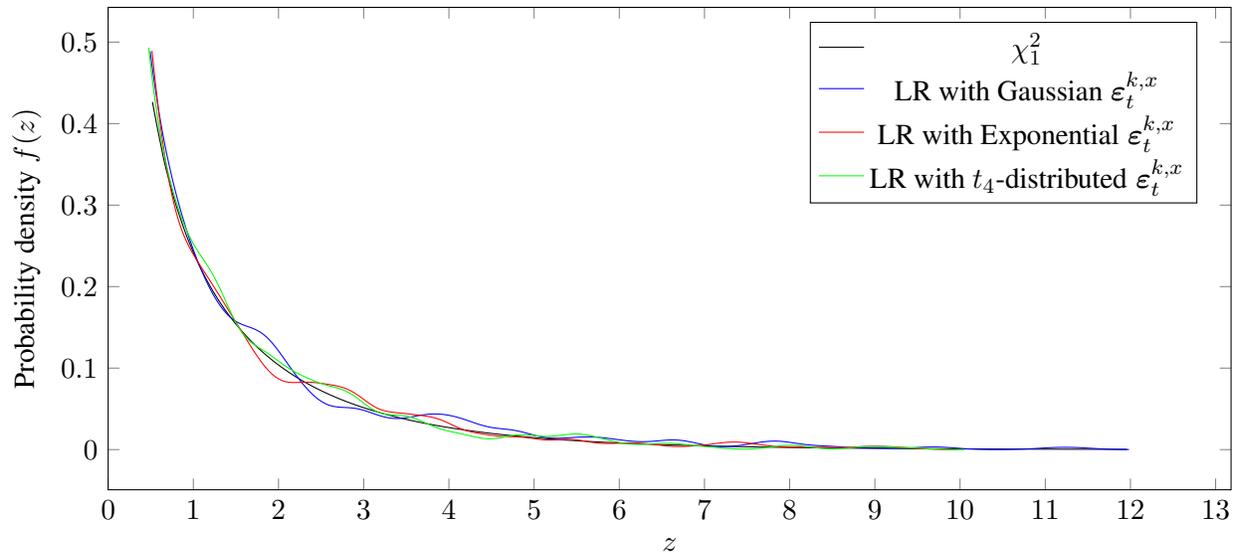
\begin{figure}[h!]
\centering
\begin{tikzpicture}
  \begin{axis}[%
    xlabel = $z$, width=\linewidth, height=8cm, xmin=0,
    ylabel = {Probability density $f(z)$}, table/col sep=comma,
    samples = 200,
    restrict y to domain = 0:0.5, legend pos=north east,
     restrict x to domain = 0.01:12]
    \addplot+[black, smooth, no marks] table[x index=0,y index=1] {chisq_density.csv}; \addlegendentry{$\chi^2_1$}
    \addplot+[blue, smooth, no marks] table[x index=0,y index=1] {LR_density.csv}; \addlegendentry{LR with Gaussian $\vvarepsilon_t^{k,x}$}
    \addplot+[red, smooth, no marks] table[x index=2,y index=3] {LR_density.csv}; \addlegendentry{LR with Exponential $\vvarepsilon_t^{k,x}$}
    \addplot+[green, smooth, no marks] table[x index=4,y index=5] {LR_density.csv}; \addlegendentry{LR with $t_4$-distributed $\vvarepsilon_t^{k,x}$}
  \end{axis}
\end{tikzpicture}
\caption{Probability densities of the LR tests obtained under the null hypothesis that $\rho=0$, for $n_{\text{sim}}=1000$ and for the three model specifications discussed at the beginning of Appendix \ref{simulations appendix}, respectively, a Gaussian, an Exponential, and a Student's $t_4$ distribution for the idiosyncratic components, $\vvarepsilon^{k,x}_t$. We compare these probability densities to a $\chi^2_1$ distribution, which is the asymptotic distribution of the LR test under the null hypothesis, and a correct specification of the model.}
\label{fig:LR_distribution}
\end{figure}

We finally look at the consequences of non-Gaussian idiosyncratic components, on the finite-sample distribution of the LR test for the null hypothesis that $\rho=0$. The formula for computing the LR test is $LR = -2(\mathcal{L}_R - \mathcal{L})$, where $\mathcal{L}_R$ is the value of the log-likelihood under the restriction imposed by the null hypothesis, and $\mathcal{L}$ is the value of the log-likelihood for the unrestricted model, which estimates $\rho$. We simulate data for each of the three model specifications discussed at the beginning of this section with $\rho=0$, and we calculate the values of the LR test. We do this for 1000 simulation runs. Notice that under the null hypothesis that $\rho=0$, and a correct specification of the model, the LR test should be asymptotically $\chi^2_1$-distributed, which (as mentioned in Section \ref{empirics}) does not necessarily hold if the model is misspeficied, e.g. if the true distribution of the error terms is not Gaussian, but a Gaussian distribution is instead used in order to estimate the model. In Figure \ref{fig:LR_distribution} we therefore compare the probability densities of the LR tests obtained as described above, to a $\chi^2_1$ distribution. We notice that the density of the LR test is not sensitive to deviations of the idiosyncratic components from Gaussianity. For all three distributions of the error terms considered, i.e., Gaussian, Exponential, and Student's $t$ with 4 degrees of freedom, the density of the LR test is close to a $\chi^2_1$ distribution. These simulation results suggest that our method allows to conduct inference as usual based on the results of the LR test, even if the distribution of the idiosyncratic components is misspecified.

\begin{table}[h!]
\centering
\medskip
\begin{tabular}{lrrrrrrr} \toprule
      & \multicolumn{1}{c}{$\rho=0$} & \multicolumn{1}{c}{$\rho=0.2$} & \multicolumn{1}{c}{$\rho=0.4$} & \multicolumn{1}{c}{$\rho=0.6$} & \multicolumn{1}{c}{$\rho=0.8$} & \multicolumn{1}{c}{$\rho=0.9$} & \multicolumn{1}{c}{$\rho=0.99$}          \\ \midrule 
         &   \multicolumn{7}{c}{Gaussian-distributed idiosyncratic components} \\ \cmidrule(lr){2-8}         
  $\MSFE(\hat{L}^k_{t|\Omega_t^{-}})$        & 1.016 & 0.988 & 0.994 & 0.941 & 0.890 & 0.835 & 0.773 \\
  $\var(\hat{L}^k_{t|\Omega_t^{-}})$         & 1.016 & 0.988 & 0.994 & 0.941 & 0.890 & 0.835 & 0.773 \\
  $\bias^2(\hat{L}^k_{t|\Omega_t^{-}})$        & 0.945 & 3.805 & 0.992 & 0.818 & 0.608 & 1.302 & 1.040 \\
 $\MSFE(\hat{R}^k_{t|\Omega_t^{-}})$         & 1.048 & 0.982 & 0.924 & 0.754 & 0.580 & 0.411 & 0.253 \\
  $\var(\hat{R}^k_{t|\Omega_t^{-}})$        & 1.047 & 0.981 & 0.924 & 0.754 & 0.580 & 0.411 & 0.253 \\
    $\bias^2(\hat{R}^k_{t|\Omega_t^{-}})$      &1.266 & 2.152 & 0.884 & 2.102 & 0.795 & 0.125 & 0.663 \\ \midrule
   &   \multicolumn{7}{c}{Exponentially-distributed idiosyncratic components} \\ \cmidrule(lr){2-8} 
$\MSFE(\hat{L}^k_{t|\Omega_t^{-}})$ & 1.012 & 1.004 & 0.985 & 0.923 & 0.880 & 0.864 & 0.804 \\
 $\var(\hat{L}^k_{t|\Omega_t^{-}})$ &   1.012 & 1.003 & 0.985 & 0.923 & 0.880 & 0.864 & 0.804 \\
 $\bias^2(\hat{L}^k_{t|\Omega_t^{-}})$ &   0.886 & 1.217 & 0.124 & 1.033 & 551.692 & 0.267 & 0.498 \\
 $\MSFE(\hat{R}^k_{t|\Omega_t^{-}})$ &   1.041 & 0.998 & 0.947 & 0.754 & 0.547 & 0.434 & 0.286 \\
 $\var(\hat{R}^k_{t|\Omega_t^{-}})$ &   1.042 & 0.995 & 0.946 & 0.749 & 0.538 & 0.429 & 0.276 \\
 $\bias^2(\hat{R}^k_{t|\Omega_t^{-}})$ &  0.172 & 1.774 & 2.579 & 6.216 & 390.045 & 17.565 & 12.570 \\  \midrule
    &   \multicolumn{7}{c}{t-distributed idiosyncratic components} \\ \cmidrule(lr){2-8}
$\MSFE(\hat{L}^k_{t|\Omega_t^{-}})$          &  1.011 & 1.017 & 0.992 & 0.970 & 0.935 & 0.815 & 0.788 \\
$\var(\hat{L}^k_{t|\Omega_t^{-}})$          &  1.012 & 1.017 & 0.992 & 0.970 & 0.934 & 0.815 & 0.788 \\
$\bias^2(\hat{L}^k_{t|\Omega_t^{-}})$          &  0.927 & 1.233 & 0.869 & 13.867 & 2.116 & 3.330 & 7.463 \\
$\MSFE(\hat{R}^k_{t|\Omega_t^{-}})$          & 1.039 & 1.015 & 0.937 & 0.782 & 0.564 & 0.381 & 0.231 \\
$\var(\hat{R}^k_{t|\Omega_t^{-}})$          & 1.039 & 1.015 & 0.937 & 0.782 & 0.564 & 0.380 & 0.231 \\
$\bias^2(\hat{R}^k_{t|\Omega_t^{-}})$          &  0.965 & 4.126 & 0.091 & 0.840 & 1.119 & 90.106 & 53.658 \\
\bottomrule
\end{tabular}
\caption{Simulation results from the three settings described in Appendix \ref{simulations appendix}. The values are reported relative to the respective measures calculated from the model that does not include the auxiliary series; values $<1$ are in favour of our method.  $n_{\text{sim}}=1000$.}
\label{tab:simulations_NG}
\end{table}

\newpage

\section{Additional empirical results} \label{appendix_empirics}

\begin{figure}[H]
\centering
\begin{tikzpicture}
\begin{axis}[xbar,xmin=0.4,xmax=1,bar width=5pt,table/col sep=comma, ytick=data, axis lines*=left, enlarge y limits=0.01,
yticklabels from table={freq_monthly_EN.csv}{name},width=0.8\textwidth, height=0.85\textheight,xlabel=Frequency,yticklabel style = {font=\scriptsize}]
    \addplot[fill=gray] table[y expr=-\coordindex, x=freq] {freq_monthly_EN.csv};   
  \end{axis}
\end{tikzpicture}
\caption{Frequency of monthly Google search terms selection by the elastic net in the out-of-sample period. A value of 1 means that the variable has been selected in every month of the out-of-sample period. We only report search terms that have been selected at least 50\% of the times.}
\label{fig:freq_monthly}
\end{figure}

\begin{figure}[H]
\centering
\begin{tikzpicture}
\begin{axis}[xbar,xmin=0.4,xmax=1,bar width=5pt,table/col sep=comma, ytick=data, axis lines*=left, enlarge y limits=0.01,
yticklabels from table={freq_agg_EN.csv}{name},width=0.8\textwidth, height=0.85\textheight,xlabel=Frequency,yticklabel style = {font=\scriptsize}]
    \addplot[fill=gray] table[y expr=-\coordindex, x=freq] {freq_agg_EN.csv};   
  \end{axis}
\end{tikzpicture}
\caption{Frequency of weekly Google search terms (aggregated to the monthly frequency according to equation \eqref{eq:aggregation}) selection by the elastic net in the out-of-sample period. A value of 1 means that the variable has been selected in every week of the out-of-sample period. We only report search terms that have been selected at least 50\% of the times.}
\label{fig:freq_agg}
\end{figure}
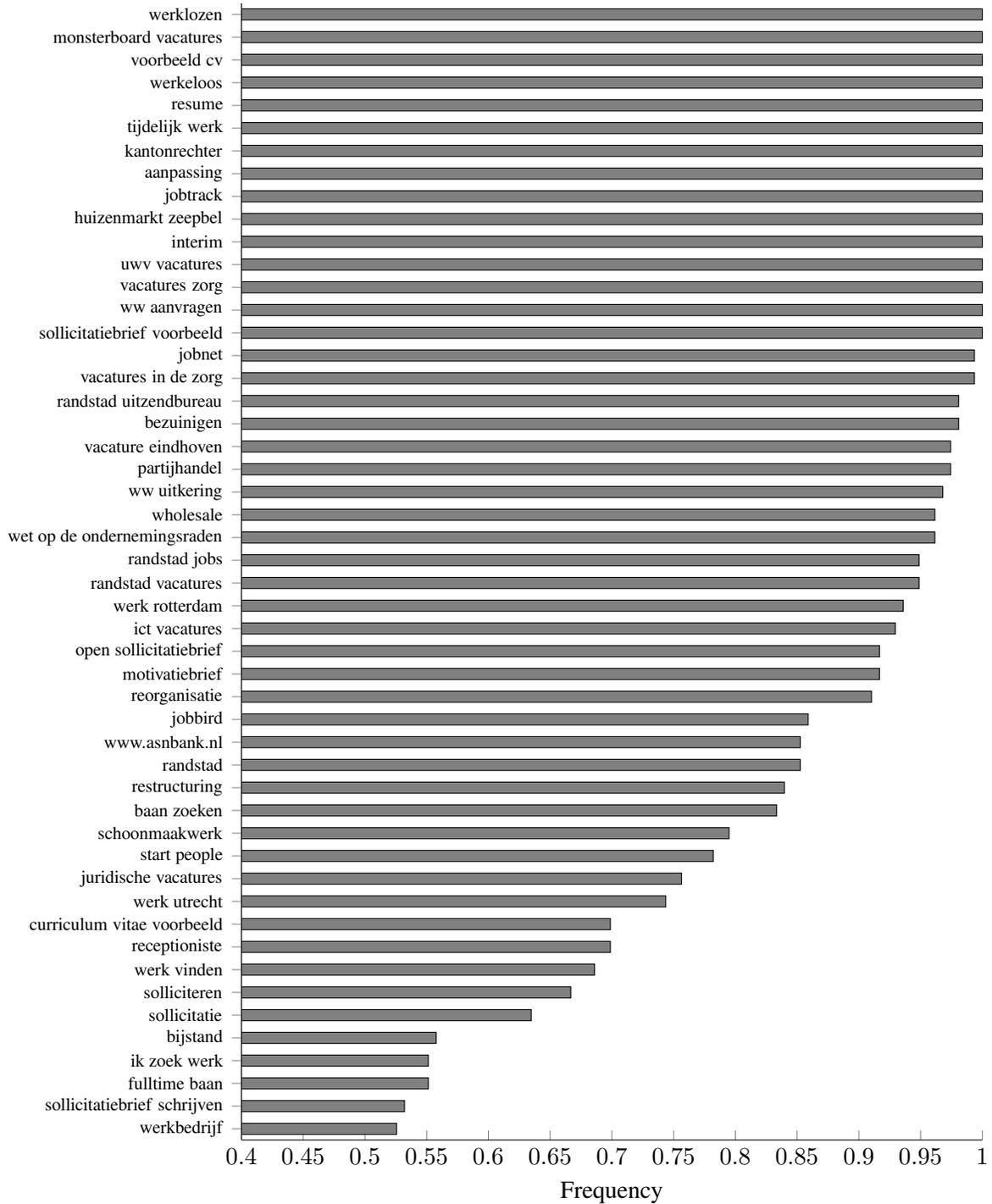


\begin{figure}[H]
\centering
\begin{tikzpicture}
\begin{axis}[ybar=-5pt, width=0.9\linewidth, height=6cm, axis lines*=left, ylabel={p-value},
    bar width=5pt, table/col sep=comma, legend pos=north west,  legend columns=2, 
    xtick=\empty 
    ]
        \addplot[fill=white, select coords between index={0}{4}] table[y=BS, x=n]{pnorm_monthly.csv}; \addlegendentry{$\tilde{\vv}_t^{k,y}$}
        \addplot[fill=gray, select coords between index={5}{5}] table[y=BS, x=n]{pnorm_monthly.csv}; \addlegendentry{$\tilde{v}_t^{k,CC}$}
        \addplot[fill=black, select coords between index={6}{44}] table[y=BS, x=n]{pnorm_monthly.csv}; \addlegendentry{$\tilde{\vv}_t^{k,GT}$}
        \addplot[red,line legend,sharp plot] coordinates {(1,0.05) (45,0.05)}; \addlegendentry{0.05}
\end{axis}
\end{tikzpicture}
\caption{p-values from the Bowman-Shenton test for individual normality, performed on each of the standardized prediction errors of the labour force, the claimant counts, and the Google Trends series ($\tilde{\vv}^{k}_t$). The standardized prediction errors are obtained from the CC \& GT model which employs the monthly Google Trends and include two of their factors. The red line represents the confidence level of 0.05.}
\label{fig:pvalues_norm_monthly_BS}
\end{figure}

\begin{figure}[H]
\centering
\begin{tikzpicture}
\begin{axis}[ybar=-5pt, width=0.9\linewidth, height=6cm, axis lines*=left, ylabel={p-value},
    bar width=5pt, table/col sep=comma, legend pos=north east,  legend columns=2, 
    xtick=\empty 
    ]
        \addplot[fill=white, select coords between index={0}{4}] table[y=BS, x=n]{pnorm_weekly.csv}; \addlegendentry{$\tilde{\vv}_t^{k,y}$}
        \addplot[fill=gray, select coords between index={5}{5}] table[y=BS, x=n]{pnorm_weekly.csv}; \addlegendentry{$\tilde{v}_t^{k,CC}$}
        \addplot[fill=black, select coords between index={6}{42}] table[y=BS, x=n]{pnorm_weekly.csv}; \addlegendentry{$\tilde{\vv}_t^{k,GT}$}
        \addplot[red,line legend,sharp plot] coordinates {(1,0.05) (43,0.05)}; \addlegendentry{0.05}
\end{axis}
\end{tikzpicture}
\caption{p-values from the Bowman-Shenton test for individual normality, performed on each of the standardized prediction errors of the labour force, the claimant counts, and the Google Trends series ($\tilde{\vv}^{k}_t$). The standardized prediction errors are obtained from the CC \& GT model which employs the weekly Google Trends and include two of their factors, and which iterates the estimation of $\mLambda$ and $\mPsi$. The red line represents the confidence level of 0.05.}
\label{fig:pvalues_norm_weekly_BS}
\end{figure}


\end{document}